\documentclass[aps,prx,twocolumn,superscriptaddress,floatfix]{revtex4-2}

\usepackage{graphicx}
\usepackage{xcolor}
\usepackage{dcolumn}
\usepackage{physics}
\usepackage{hyperref}
\hypersetup{
  colorlinks=true,
  linkcolor=blue,   
  citecolor=blue,   
  urlcolor=blue,    
}
\usepackage[clock]{ifsym}

\usepackage{bm}
\usepackage{multirow}
\usepackage{booktabs}
\usepackage{amssymb}
\usepackage{longtable}
\usepackage{array}
\usepackage{orcidlink}
\usepackage{braket}

\newcommand{\figref}[2]{\hyperref[#1]{Fig.~\ref*{#1}#2}}

\makeatletter
\let\switch@array\relax
\makeatother

\newcommand{\clk}{\raisebox{-0.1ex}{\scalebox{0.75}{\VarClock}}}  
\begin{document}

\preprint{APS/123-QED}

\title{A Platform-aware Compilation Framework for Fault-tolerant Quantum Computation}

\author{Srushti Patil\,\orcidlink{0000-0001-6853-0604}}
\email{srushti.patil@nbi.ku.dk}
\affiliation{NNF Quantum Computing Programme, Niels Bohr Institute, University of Copenhagen, Blegdamsvej 17, DK-2100 Copenhagen Ø, Denmark}

\author{Susan X. Chen\,\orcidlink{0000-0003-2049-8558}}
\affiliation{Quantum Engineering Centre for Doctoral Training, H. H. Wills Physics Laboratory and School of Electrical, Electronic, and Mechanical Engineering, University of Bristol, BS8 1FD, United Kingdom}
\affiliation{NNF Quantum Computing Programme, Niels Bohr Institute, University of Copenhagen, Blegdamsvej 17, DK-2100 Copenhagen Ø, Denmark}

\author{Andreas Juul Bay-Smidt\,\orcidlink{0009-0005-9428-6664}}
\affiliation{NNF Quantum Computing Programme, Niels Bohr Institute, University of Copenhagen, Blegdamsvej 17, DK-2100 Copenhagen Ø, Denmark}
\affiliation{Nano-Science Center and Department of Chemistry, University of Copenhagen, Denmark}

\author{Stefan Alaric Schäffer\,\orcidlink{0000-0002-5296-0332}}
\affiliation{NNF Quantum Computing Programme, Niels Bohr Institute, University of Copenhagen, Blegdamsvej 17, DK-2100 Copenhagen Ø, Denmark}

\author{Peter Krogstrup\,\orcidlink{0000-0002-1930-8553}}
\affiliation{NNF Quantum Computing Programme, Niels Bohr Institute, University of Copenhagen, Blegdamsvej 17, DK-2100 Copenhagen Ø, Denmark}

\author{Stefano Paesani\,\orcidlink{0000-0001-5709-0906}}
\email{stefano.paesani@nbi.ku.dk} 
\affiliation{NNF Quantum Computing Programme, Niels Bohr Institute, University of Copenhagen, Blegdamsvej 17, DK-2100 Copenhagen Ø, Denmark}

\author{Gemma C. Solomon\,\orcidlink{0000-0002-2018-1529}}
\email{gsolomon@chem.ku.dk} 
\affiliation{NNF Quantum Computing Programme, Niels Bohr Institute, University of Copenhagen, Blegdamsvej 17, DK-2100 Copenhagen Ø, Denmark}
\affiliation{Nano-Science Center and Department of Chemistry, University of Copenhagen, Denmark}

\date{\today}

\begin{abstract}
The compilation of an algorithm can vary significantly with the choice of physical hardware platform and error correction model.
Yet, current compilation frameworks typically commit to a single architecture-hardware configuration, making it difficult to assess resource estimates across platforms.
We present a platform-aware compilation framework that re-compiles a quantum circuit into a hardware-compatible instruction set as well as fault-tolerant operations and provides end-to-end resource estimates in terms of physical-qubit count, time-to-solution, and classical processing time.
We benchmark the framework by obtaining end-to-end resource estimates for different compilers, each tailored to the functionalities of specific hardware modalities: connectivity, clock speed, and noise model.
As part of this framework, we introduce a transversal active volume (t-AV) compilation architecture designed for the efficient execution of fault-tolerant operations in platforms supporting long-range logical connectivity. 
We benchmark the framework for Hamiltonian simulation of the 2D Fermi-Hubbard model as well as for eigenenergy estimation of a small molecule (trimethylenemethane) as a candidate for early fault-tolerant demonstration of quantum chemistry.
For the latter, we show that end-to-end quantum simulations can be achieved with $\sim10^4$ physical qubits and runtimes ranging from $10^2$ ms (photonics, superconducting) to $10^5$ ms (neutral atoms). 
\end{abstract}

\maketitle

\section{Introduction}
\label{sec:introduction}
As quantum technologies move towards early fault-tolerant quantum computing (FTQC), the community has advanced efforts to shift away from asymptotic scalings~\cite{Shor_1997,grover1996fastquantummechanicalalgorithm, HamiltonianSimulationAsymph} toward concrete resource estimates in terms of logical qubit count and T/Toffoli gate counts~\cite{REShorsEllipticCurve, RENearTermQChemVQE, REShorsLowResource, REShorsWith2n+3Qubits, REGateCountSmallQComp, REDrugDiscovery, RENearTermhubbardVQE}.
To capture the full cost of a fault-tolerant implementation, refined resource estimation pipelines have emerged that consider the underlying quantum error correction (QEC) scheme, classical processing time, and a pre-assumed layout~\cite{EvenMoreEfficient, REquantumCSAlgorithms, QREChemTheory, REAzure}.
These full-stack estimations provide a more realistic picture of the implementation costs but often remain agnostic to the capabilities of specific hardware platforms.

Hardware-aware compilation architectures, however, can recompile a quantum circuit using a set of operations that are compatible with a given QEC scheme and amenable to the platform's capabilities and constraints~\cite{PBCGameSurface, AVLitinski, ArchitectNeutralAtoms, PBCqLDPC, ArchitectqBBC, REforIonTrapsWithLS, ArchitectNeutralAtoms, hardwareTailoredRE, AVFasterQChem}.
For instance, a circuit initially compiled for architectures limited by nearest-neighbor connectivity and high spatial routing overhead~\cite{PBCGameSurface} can be recompiled into an instruction set that directly exploits features of platforms supporting limited non-local connectivity~\cite{AVLitinski, AVFasterQChem, apel2026compiling2dfermihubbardgroundstate}.
Moreover, if a platform supports all-to-all connectivity, the same algorithm can be compiled more efficiently with constant-depth unitaries, given a hardware-compatible QEC scheme~\cite{ArchitectNeutralAtoms, khan2026architectingearlyfaulttolerant, LowOverheadTransversal, LowOverheadTransversalArchitecture}.
However, because existing compilation architectures commit to a specific platform, they cannot be directly translated to another platform without incurring substantial space-time overhead and consequently suboptimal physical circuits.
This non-transferability makes it difficult to assess the near- or long-term nature of a given use case, as resulting resource estimates can change by orders of magnitude from one platform to another.
Since multiple platforms with different capabilities are emerging in parallel, in the future, one may want to choose which hardware best suits a given algorithm.  
A unified compilation framework with unified metrics that can generate accurate resource estimates across platforms is needed: both to enable fair cross-platform comparison and to help algorithm researchers choose a hardware platform suited to their requirements. 

\begin{figure*}
\centering
\includegraphics[width=1\linewidth]{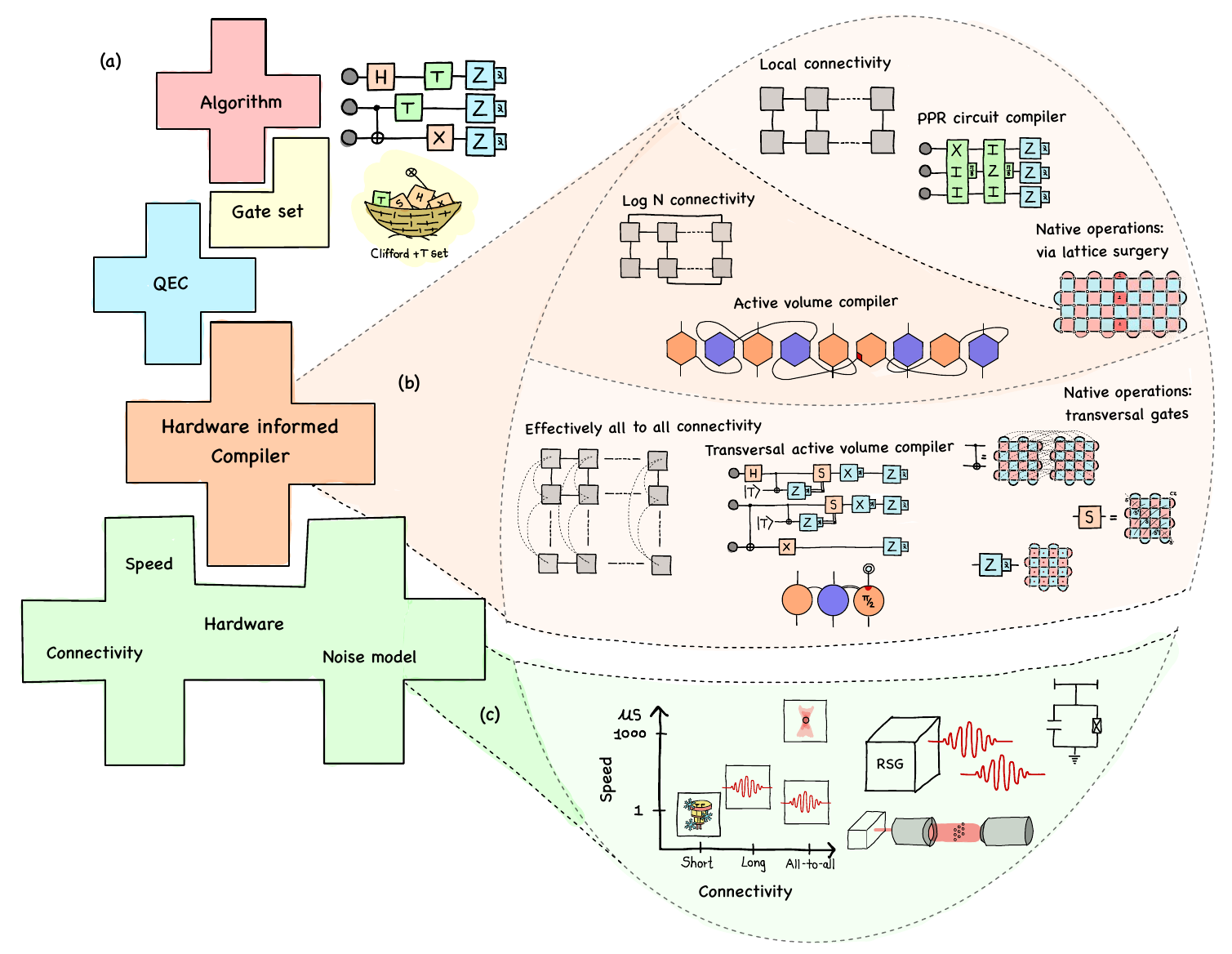}
\caption{Overview of the platform-aware compilation framework.
(a) The full FTQC pipeline is divided into an Algorithm, a QEC, and a Hardware stack.
The algorithm is expressed as a logical circuit in a universal gate set (here, Clifford$+T$), which the QEC code must support (surface code in this case). The hardware-informed compiler then bridges the QEC and Hardware stacks, selecting a compilation strategy from the logical operations available in the code and the platform's connectivity, speed, and noise.
(b) Circuit compilation for the three connectivity classes we consider. Local (nearest-neighbor) connectivity: the PPR compiler rewrites the circuit as Pauli product rotations, which are executed via lattice surgery. Limited non-local ($\log N$) connectivity: the active-volume compiler exploits long-range links to reduce spacetime cost. Effectively all-to-all connectivity: the transversal active-volume compiler implements logic through transversal gates and quantifies space-time volume in relevant units. 
(c) Sketch of the different hardware modalities considered in the framework: superconducting, photonic, and neutral-atom, where we show their characteristic operation cycle time and connectivity (short, long, all-to-all). Note that we consider photonics hardware with both long ($\log N$) and all-to-all connectivity. }
\label{fig: workflow}
\end{figure*}

In this work, we introduce a platform-aware compilation framework that unifies fault-tolerant compilation architectures suited for platforms with different connectivity constraints within a single end-to-end resource-estimation pipeline.
Given a quantum algorithm expressed as a logical circuit compiled into a universal gate set, the pipeline compiles a hardware-compatible instruction set for platforms with one of the three considered connectivities: nearest-neighbor (e.g., superconducting or semiconductor spins), limited non-local (e.g., photonic), and effectively all-to-all (e.g., neutral atoms).
Using the instruction set, it then provides concrete resource estimates in terms of physical-qubit count, runtime, bridge-qubit demand for parallelization, and classical processing overhead in terms of reaction depth. 

On top of our platform-aware compilation scheme, we introduce a novel transversal active-volume (t-AV) architecture designed specifically for platforms with effectively all-to-all logical connectivity.
Within the t-AV architecture, we compile transversal logical operations into their $\mathrm{ZX}$-diagram representations \cite{Coecke_2011}, from which we calculate the space-time volume and resource estimates. 
The ZX-diagrams are useful for efficiently scheduling transversal logical operations in space-time and comparing resource estimates across platforms.
We further derive the t-AV implementation and resource costs of magic state factories - three based on distillation and one on cultivation - and optimize their spatial layout within the t-AV architecture.
We additionally combine magic state distillation factories with zero-level distillation via code conversion~\cite{ZeroLevelDistillation} to obtain $(0+1)$-level variants that substantially reduce the space-time overhead and provide an alternative to concatenated distillation on early fault-tolerant devices.

We benchmark the resource estimation pipeline using existing fault-tolerant architectures (baseline, compact, and active volume) as well as our t-AV architecture, each combined with suitable hardware platforms, on two use cases: (i) Trotterized phase estimation of the low-lying eigenenergies of a trial molecule (TMM).
(ii) Trotterized time evolution of a $10 \times 10$ square lattice Fermi–Hubbard (FH) model (200-qubit problem) as a quantum-advantage benchmark~\cite{khan2026architectingearlyfaulttolerant}. 
We find that TMM requires $\sim 10^4$ physical qubits with runtime ranging from 10-100 ms on photonics and superconducting platforms, to $10^5$ ms on the neutral atoms platform. 
The FH benchmark computation can be executed with slightly increased runtime and requires $\sim 10^5$ to $\sim 10^6$ physical qubits.  
Finally, we perform reaction depth analysis on both benchmarks and find that the runtime overhead from feedforward-limited measurement is negligible for neutral-atom t-AV and modest for photonic t-AV.

The paper is organized as follows: Sec.~\ref{sec:platform-aware-compilation-and-metrics-of-interest} introduces our platform-aware compilation framework and associated resource metrics. 
Sec.~\ref{subsec:pauli-based-computing} and~\ref{subsec:t-clifford-t-computation} review universal computation models: Pauli-based computation and gate-based transversal logic, respectively. 
In Sec.~\ref{subsec:cost-assignment-for-native-operations}, we assign a time cost to each native operation, followed by a discussion on non-Clifford resources in Sec.~\ref{subsec:non-clifford-resources}.
We then outline compilation architectures for local and limited non-local connectivity platforms in Sec.~\ref{subsec:quantum-computer-with-local-connectivity} and~\ref{subsec:quantum-computer-with-limited-non-local-connectivity} respectively. 
In Sec.~\ref{subsec:transversal-active-volume-architecture}, we introduce our t-AV architecture for platforms supporting effectively all-to-all connectivity and transversal Cliffords. 
Finally, Sec.~\ref{sec:results} presents the end-to-end resource estimates and cross-platform comparisons.


\section{Platform aware compilation and metrics of interest}
\label{sec:platform-aware-compilation-and-metrics-of-interest}

Resource estimates depend on the compatibility of the logical operations with both the chosen QEC scheme and the underlying hardware platform.
To make this dependence explicit, we divide our compilation framework into the three stacks shown in Fig.~\ref{fig: workflow}\textcolor{blue}{(a)}: An algorithm stack, a QEC stack, and a Hardware stack. 
The algorithm and QEC stacks are connected by expressing an algorithm as a logical circuit compiled into a universal gate set, here taken to be Clifford+$T$. 
This gate set specifies the logical operations that the QEC scheme must be able to execute via a set of supported native operations.
The QEC and Hardware stacks are, in turn, connected through our hardware-informed compiler, shown in Fig.~\ref{fig: workflow}\textcolor{blue}{(b)}. 
Our compiler is divided into three parts depending on the connectivity of the hardware. 
Throughout this paper, we focus on the following three connectivity classes (with $N =$ number of logical qubits): 
    \begin{itemize}
        \item \textit{Nearest-neighbour (local)}: a physical qubit is connected only to neighboring physical qubits. Typically realized in superconducting and semiconductor platforms.
        \item $\mathit{\log N}$: each physical qubit of a logical patch couples to $\log N$ physical qubits residing on distinct logical patches, e.g., photonics/spin qubit platforms, where physical connections facilitate logical connectivity between two patches.  
        \item \textit{(Effectively) All-to-all}: each physical qubit of a logical patch can couple to $\sim N$ physical qubits on the remaining logical patches. This is realizable in, for instance, neutral atoms and photonics platforms.
    \end{itemize}
The most efficient compilation of a logical circuit depends on the connectivity of the hardware.  
If the platform is limited to local connectivity between physical qubits and a planar grid layout, where long-range logical/physical operations cannot be executed directly, the logical circuit as well as the qubit layout must be reformulated such that it respects this locality constraint.
Pauli-based computation (PBC) and lattice surgery offer one such route where logical operations are performed via joint Pauli product measurements (PPMs), by measuring stabilizers along the shared boundary of, for example, adjacent surface-code patches~\cite{PBCGameSurface}.
The circuit is rewritten in terms of multi-qubit $\pi/8$ Pauli product rotations (PPRs) and executed via PPMs (see Sec. \ref{subsec:pauli-based-computing}).
We provide more detailed discussions on PPRs/PPMs in Appendix~\ref{app:pauli-based-computation}.
If the platform supports $\log \mathrm{N}$ non-local physical connections per physical component of a logical patch, $\mathrm{N}$ being the total number of logical qubits, the same PPR circuit can be recompiled into a network of \textit{logical blocks}, which is a $\mathrm{ZX}$ representation with a specified orientation~\cite{AVLitinski}.
Sec.~\ref{subsec:quantum-computer-with-limited-non-local-connectivity} provides a brief description of the construction of logical blocks.
An outline of architectures that use such constructions is given in Appendix~\ref{app:architectures}.
If the platform instead supports effectively all-to-all connectivity ($\approx \mathrm{N}$ connections per physical component) and transversal Cliffords, and therefore is compatible with the t-AV architecture introduced in this paper, one can recompile the circuit into transversal logical blocks derived from $\mathrm{ZX}$ circuit diagrams, as described in Sec.~\ref{subsec:transversal-active-volume-architecture}.
The hardware-informed circuit compiler, therefore, takes into account both the logical operations supported by the QEC scheme and the constraints imposed by the hardware, such as connectivity and gate speed, illustrated in Fig.~\ref{fig: workflow}\textcolor{blue}{(c)}, as well as the noise model, and re-compiles the circuit into a hardware-compatible fault-tolerant instruction set.
Our framework can, in general, be extended to error-correcting codes supporting measurement-based logical operations or transversal Cliffords (e.g., dynamic surface codes \cite{eickbusch2025demonstratingdynamicsurfacecodes}, or qLDPC codes \cite{qldpcPPMs, PBCqLDPC}); however, for simplicity, in this study we consider rotated surface codes as an underlying error correction scheme. 

Throughout the paper, we represent a surface code patch encoding a logical qubit as a $(d_X, d_Z, d_m)$ cube, where $d_X$ and $d_Z$ are spatial code distances associated with the $X$ (red) and $Z$ (blue) boundaries, respectively, and $d_m$ is the number of syndrome-measurement rounds, which plays the role of a temporal code distance.
Every logical operation that we do with surface codes can be carved into $(d_X, d_Z, d_m)$ cubes with specific boundary conditions.
Unless otherwise specified, we assume $d_X = d_Z = d_m = d$ in a PBC/lattice-surgery-based computation.
In computation with t-Cliffords, $d_m \to O(1)$, assuming fast transversal decoders~\cite{scalableDecoders, CorrelatedDecoding, LowOverheadTransversal}. 
Below, we outline further useful notation, conventions, and resource metrics used throughout the paper and include a list of abbreviations and symbols in Table~\ref{tab:abbreviations}.
\begin{itemize}
    \item  \textit{Runtime}: The total number of logical clock cycles required to complete the computation, including distillation and feedforward overheads. One \textit{logical clock cycle} corresponds to $d$ rounds of syndrome extraction, where $d$ is the surface code distance, and each round of syndrome extraction is referred to as a \textit{code cycle}.
    We denote a logical clock cycle as $1 \clk$, and a code cycle as $1 \clk_c$, with $1 \clk = d \clk_c$.
    \item \textit{Physical qubit count:} The total number of physical qubits required to execute a given computation, including data, ancilla, bridge, and factory resources. Each logical qubit is encoded in a rotated surface code patch/tile, which requires $2d^2$ physical qubits, including measurement ancillas. We denote a single code patch by $\boxdot$.
    \item \textit{Space-time volume (STV)}: The product of runtime (in terms of code cycles) and logical qubit count, which provides a measure of the overall implementation cost. Depending on the architecture and hardware capabilities, the space-time volume can be quantified using only the `active' part of the computation, referred to as the \textit{active volume}. For the t-AV architecture developed in this work, we calculate the \textit{transversal active volume} of each operation, which is a space-time volume of active logical operations, implemented transversely.
    Whenever the logical qubit count is converted to physical qubits via the surface-code encoding ($2d^2$ physical qubits per logical patch), we refer to the resulting quantity as the \textit{physical STV}: the runtime in code cycles multiplied by the physical-qubit count.
    \item \textit{Reaction time and depth:} The non-Clifford operations need to be applied via magic state injection, which requires Clifford corrections. Such Clifford corrections can be implemented via adaptively chosen basis measurements. 
    A measurement whose basis depends on prior measurement outcomes is called a reactive measurement, and such measurements cannot be deferred to the end and must be executed in place. The time required to classically process the outcome of a previous measurement and update the basis of the next measurement accordingly is called reaction time $\tau_r$.
    The number of such sequential basis-update layers in any given logical cycle is called the reaction depth $R$, and this is used to quantify the classical latency introduced due to these feedforward-limited measurements.
\end{itemize}

\begin{table}[t!]
\vspace*{-0.7\baselineskip}
\centering
\caption{Abbreviations and symbols used throughout this work, with the section where each is introduced.}
\label{tab:abbreviations}
\resizebox{\columnwidth}{!}{
\begin{tabular}{l l c}
\hline\hline
 & Meaning & Referenced Section \\
\hline
\multicolumn{3}{l}{\textit{Abbreviations}} \\
QEC     & Quantum error correction & Sec.~\ref{sec:introduction} \\
FTQC    & Fault-tolerant quantum computing & Sec.~\ref{sec:introduction} \\
PBC     & Pauli-based computation & Sec.~\ref{subsec:pauli-based-computing} \\
PPR     & Pauli product rotation & Sec.~\ref{subsec:pauli-based-computing} \\
PPM     & Pauli product measurement & Sec.~\ref{subsec:pauli-based-computing} \\
t-Clifford & Transversal logical Clifford & Sec.~\ref{subsec:t-clifford-t-computation} \\
MSD     & Magic state distillation & Sec.~\ref{subsec:non-clifford-resources} \\
AV      & Active volume & Sec.~\ref{subsec:quantum-computer-with-limited-non-local-connectivity} \\
t-AV    & Transversal active volume & Sec.~\ref{subsec:transversal-active-volume-architecture} \\
STV     & Space-time volume & Sec.~\ref{sec:platform-aware-compilation-and-metrics-of-interest} \\
tauble  & Transversal logical block & Sec.~\ref{subsec:transversal-active-volume-architecture} \\
Trans-dist  & Transversal 15-to-1 factory & Sec.~\ref{subsubsec:transversal-15-to-1-distillation} \\
Parity-dist & Parity-ancilla 15-to-1 factory & Sec.~\ref{subsubsec:distillation-with-parity-ancilla} \\
LS-dist     & Lattice-surgery factory & Sec.~\ref{subsubsec:lattice-surgery-based-distillation-factories} \\
0-dist      & Zero-level distillation & Sec.~\ref{subsubsec:concatenation-of-msd-protocols-with-zero-level-distillation} \\
Cult        & Fold-transversal cultivation & Sec.~\ref{subsubsec:fold-transversal-cultivation} \\
TMM     & Trimethylenemethane & Sec.~\ref{subapp:trotterized-quantum-phase-estimation} \\
PPP     & Pariser--Parr--Pople model & Sec.~\ref{subapp:trotterized-quantum-phase-estimation} \\
QPE & Quantum phase estimation & Sec.~\ref{sec:results} \\
QPE-Abs & QPE for absolute energies & Sec.~\ref{sec:results} \\
Stat-QPE(-Gap) & Statistical QPE (for the energy gap) & Sec.~\ref{sec:results} \\
\hline
\multicolumn{3}{l}{\textit{Symbols}} \\
$d$ ($d_X, d_Z, d_m$) & Code distances (spatial, temporal) & Sec.~\ref{sec:platform-aware-compilation-and-metrics-of-interest} \\
$1\,\clk = d\,\clk_c$ & Logical clock cycle $=$ $d$ code cycles & Sec.~\ref{sec:platform-aware-compilation-and-metrics-of-interest} \\
$\boxdot$ & One surface-code patch ($2d^2$ physical qubits) & Sec.~\ref{sec:platform-aware-compilation-and-metrics-of-interest} \\
$n_T$ & Total $T$-gate count of a benchmark & Sec.~\ref{subsec:non-clifford-resources} \\
$p$, $p_{out}$ & Physical / distilled magic-state error rate & Sec.~\ref{subsec:non-clifford-resources} \\
$p_L$, $\epsilon$ & Logical error rate; failure budget & Sec.~\ref{subapp:noise-model-and-distance-calculation} \\
$n_Q$, $n_C$ & Logical tiles and cycles of a computation & Sec.~\ref{subapp:noise-model-and-distance-calculation} \\
$N_{fac}$ & Factory count & Sec.~\ref{subsubsec:transversal-15-to-1-distillation} \\
$\tau_c$, $\tau_r$ & Code cycle time; reaction time & Sec.~\ref{sec:platform-aware-compilation-and-metrics-of-interest} \\
$R$, $k_{\max}$ & Reaction depth; worst-case depth per cycle & Sec.~\ref{subsec:reaction-depth-in-t-av-architecture} \\
$r$ & Number of Trotter steps & Sec.~\ref{app:quantum-simulation-problems} \\
\hline\hline
\end{tabular}
}
\end{table}

\section{Universal fault-tolerant computational models}
\label{sec:universal-computation-models}
In this section, we describe the two computational models that serve as the basis for the compilation schemes presented in the following sections. 

\subsection{Pauli-based computing}
\label{subsec:pauli-based-computing}
Pauli-based computation (PBC) is a universal computation model, well-suited to platforms with limited non-local connectivity, that allows us to express the entire computation as a sequence of $\pi/8$ PPRs instead of one- and two-qubit unitary gates. 
Logical gates can be rewritten in terms of PPRs of the form $P_{\phi} = e^{iP\phi}$, where $P$ is a multi-qubit Pauli string consisting of $X, Y, Z$ Pauli operators. 
The Hadamard can be written as $H = Z_{\pi/4}\cdot X_{\pi/4}\cdot Z_{\pi/4}$, and similarly, $S = Z_{\pi/4}$, $T = Z_{\pi/8}$ and $CNOT = (Z \otimes X)_{\pi/4} \cdot (I \otimes X)_{-\pi/4} \cdot (Z \otimes I)_{-\pi/4}$. 
With these rewrites and additional commutation rules (see Ref.~\cite{PBCGameSurface} for details), all Clifford gates can be commuted past each other to the end of the circuit and subsequently absorbed into the measurements, leaving only $\pi/8$ PPRs, $P_{\pi/8} = e^{iP\frac{\pi}{8}}$, in the circuit. 
PPRs are performed via PPMs followed by a Clifford correction, which is explained further in Appendix~\ref{app:pauli-based-computation}. 
The surface code supports multi-qubit PPMs via lattice surgery at a cost of $1 \clk$ per measurement, regardless of weight.
However, layout assumptions can further restrict the accessible boundaries, thus requiring more than one logical cycle.

\subsection{Transversal Clifford + T computation}
\label{subsec:t-clifford-t-computation}
A logical gate is \textit{transversal} if it decomposes into single- or two-qubit physical gates applied independently to each data qubit (or pair) of the code patch(es); because no single physical fault spreads onto multiple data qubits within one patch, transversal implementations are fault-tolerant by construction~\cite{FowlerSurfFTQC}.
The Eastin-Knill theorem~\cite{EastinKnillTheorem} precludes any quantum error-correcting code from admitting a universal transversal gate set. 
Hence, for a given code, only a limited number of logical gates have (fold-)transversal implementations, and for surface codes, these are the Clifford operations.
Therefore, an alternative to the Pauli-based approach is to execute these logical Cliffords directly, as \emph{transversal} physical-gate layers, rather than commuting them through the circuit.
Transversal implementation is well-suited to platforms supporting effectively all-to-all connectivity between logical patches, such as neutral-atom arrays with reconfigurable traps~\cite{ArchitectNeutralAtoms}, or with all-to-all logical connectivity within a limited number of code patches, such as certain photonic systems~\cite{AVLitinski}.
On strictly local 2D layouts, transversal operations are expensive or simply not possible.
We refer to the transversal logical Cliffords used here (up to relabeling) as \emph{t-Cliffords}.
We use the following transversal Clifford primitives for the surface code:
\begin{itemize}
    \item \textit{t-CNOT}: A logical CNOT between two surface-code patches is performed by applying physical CNOTs pairwise between corresponding data qubits~\cite{FowlerSurfFTQC}.
    \item \textit{t-Hadamard}: A logical Hadamard is transversal up to a $\pi/2$ rotation of the patch. On reconfigurable hardware, this rotation can be absorbed by relabeling the physical qubits assigned to the patch, rather than physically rotating the layout~\cite{FowlerSurfFTQC}.
    \item \textit{fold-t-$S$}: A logical phase gate $S$ is \emph{fold-transversal}---it is realized by morphing~\cite{McEwen_2023} from the rotated to the unrotated surface code~\cite{TransversalLogicalCliffordGates}, applying transversal physical gates on qubits along the main diagonal while \textit{folding} the patch about its diagonal~\cite{TransversalLogicalCliffordGates} by applying entangling operations between diagonally-symmetric qubit pairs, then morphing back to the rotated surface code.
\end{itemize}

The non-Clifford T gate is not transversal on the surface code and is implemented via magic-state injection, with high-quality $\ket{T}$ states produced by distillation (see Sec.~\ref{subsec:non-clifford-resources}).

\subsection{Cost assignment for native operations}
\label{subsec:cost-assignment-for-native-operations}
The time cost associated with a logical operation depends on the underlying model of computation, whether the execution relies on PBC or t-Clifford.
In this section, we assign a time cost to the native operations used in this paper in terms of \textit{logical clock cycles} (\clk) and code cycles ($\clk_c$), and outline the assumptions used for resource estimation. 
Across all computation models, we define a \textit{critical path} as the longest chain of sequentially dependent logical operations, and the total runtime becomes the sum of code cycles along this path; operations executed in parallel contribute no additional latency.
An operation costs $0 \, \clk$ if completed within a single code cycle and the next operation on the critical path takes at least $1 \, \clk$ (e.g., PBC).
Otherwise, an operation costs $1 \clk_c$ if both it and the subsequent operation are completed within a single code cycle (e.g., t-Cliffords).  
In PBC, logical Hadamard, phase, and CNOT gates are commuted through the circuit and absorbed into the final measurements (see Sec. \ref{subsec:pauli-based-computing}); they are therefore not charged as native operations. 
We describe the relevant native operations below and summarize their costs in Table~\ref{tab:native_op_cost}.
\begin{itemize}
    \item \textit{Logical state initialization}: In PBC, a logical $\ket{0}_L$ or $\ket{+}_L$ state can be prepared in $0 \, \clk$ by initializing all physical data qubits in the corresponding product state.
    Similarly, any arbitrary logical quantum state can be initialized at the cost of $0 \, \clk$ with some probability of getting destroyed by an undetected random Pauli error.
    In t-Clifford-based computation, the same state initialization takes $1\, \clk_c$.
    \item \textit{Destructive single-qubit measurement}: In PBC, all data qubits can be measured in the chosen basis ($X/Z$) in $0 \, \clk$. In t-Clifford-based computation, it takes $1 \, \clk_c$.
    \item \textit{Two-qubit joint Pauli measurements}: In PBC, measurements of type $P_1 \otimes P_2$ where $P_1, P_2 \in \{X, Y, Z\}$ can be performed via lattice surgery \textit{merge} and \textit{split} with $1 \, \clk$; given access to $Y$ boundaries in the case of $Y$ measurements. 
    t-Clifford-based computation does not require such measurements.
    \item \textit{Logical Hadamard gate}:
    In t-Clifford-based computation, a logical Hadamard gate is a t-Hadamard,
    and takes  $1 \, \clk_c$.
    \item \textit{Logical phase gate}: In t-Clifford based computation, a logical $S$ gate is a fold-t-S, taking  $1 \, \clk_c$.
    \item \textit{Logical CNOT gate}:
    In t-Clifford-based computation, a logical CNOT between two surface code patches is a t-CNOT, requiring $1 \, \clk_c$.
    \item \textit{Bell state initialization:} With PBC on local connectivity hardware platforms, Bell state $(\ket{00}_L + \ket{11}_L)/\sqrt{2}$ can be prepared in $1 \, \clk$ by initializing a $\ket{+}_L \otimes \ket{+}_L$ across two patches followed by joint $Z \otimes Z$ measurement between them via lattice surgery. With PBC on non-local connectivity hardware platforms, if a CNOT between two qubits can be performed transversally, then a Bell state can be generated in $0 \, \clk$ by preparing a $\ket{+}_L$ state in one patch and a $\ket{0}_L$ state in the other patch, followed by a transversal CNOT.  Similarly, in t-Clifford-based computation, the Bell state is initialized via transversal CNOTs taking $1 \, \clk_c$.
    \item \textit{SWAPs}: 
    In some architectures considered in this work, SWAP operations, other than those originating from a quantum circuit, are needed to reorganize the memory for the computation. 
    Here we describe how one can implement such SWAPs. 
    With PBC on local connectivity hardware platforms, two logical patches sitting next to each other on a 2D grid can be swapped by either routing via an ancilla patch or performing lattice surgery between them, taking $1 \, \clk$. With PBC on non-local connectivity hardware platforms $0 \, \clk$, two logical qubits can be swapped via transversal physical SWAP gates. In t-Clifford-based computation, SWAPs are implemented transversally in $1 \, \clk_c$.
   
\end{itemize}
\begin{table}[tbp]
\vspace*{-0.7\baselineskip}
\centering
\caption{Cost of native operations for PBC and transversal gate-based computation, in code cycles ($\clk_c$) and logical clock cycles ($\clk = d\clk_c$).}
\label{tab:native_op_cost}
\resizebox{\columnwidth}{!}{
\begin{tabular}{c|c|c|c}
\hline
\hline
\multirow{2}{*}{\textbf{Native operations}} & \multicolumn{2}{c|}{\textbf{PBC}} & \multirow{2}{*}{\textbf{t-Clifford all-to-all}} \\
\cline{2-3}
& \textbf{local} & \textbf{non-local} & \\
\hline
Initialize $\ket{+}_L/ \ket{0}_L$
& $0 \, \clk$
& $0 \, \clk$
& $1 \, \clk_c$ \\

Basis measurement ($Z/X$)
& $0 \, \clk$
& $0 \, \clk$
& $1 \, \clk_c$ \\

Two-qubit Pauli measurements
& $1 \, \clk$
& $1 \, \clk$
& N.A. \\

Logical Hadamard gate
& Absorbed
& Absorbed
& $1 \, \clk_c$ \\

Logical phase gate
& Absorbed
& Absorbed
& $1 \, \clk_c$ \\

Logical CNOT gate
& Absorbed
& $0 \, \clk$
& $1 \, \clk_c$ \\

Bell state initialization
& $1 \, \clk$
& $0 \, \clk$
& $1 \, \clk_c$ \\

SWAPs
& $1 \, \clk$
& $0 \, \clk$
& $1 \, \clk_c$ \\

\hline
\hline
\end{tabular}
}
\end{table}

\subsection{Non-Clifford resources}
\label{subsec:non-clifford-resources}
Universal quantum computation, whether PBC or our transversal architectures, requires non-Clifford gates. 
Non-Clifford gates such as T or Toffoli gates are not transversal on surface codes, and hence they are implemented via gate teleportation: an auxiliary magic state $\ket{T} = T\ket{+}$ is injected into a data qubit via Bell measurement followed by conditional Clifford feedforward \cite{Bravyi2005UniversalQC}.
To ensure high fidelity of $T$ gates throughout the computation, $\ket{T}$ states with sufficiently low error rates need to be produced and injected.
They can be obtained by first preparing several copies of low-quality magic states, followed by distillation or cultivation of magic states~\cite{GidneyCultivation, Bravyi2005UniversalQC}.
Many magic state distillation (MSD) protocols exhibit polynomial scaling in the output error rate, with the scaling order depending on the protocol ~\cite{BravyiHaah2012}.
The choice of distillation/cultivation protocol depends on the number of $T$ gates $(n_T)$ required by the computation.
More specifically, the protocol needs to supply magic states with an output logical error rate of $p_{out} < \epsilon/n_T$, where $\epsilon$ is the target failure rate fixed for an algorithm. 
The total $T$ counts of the benchmarks considered in this work range from $n_T \approx 1.6 \times 10^{4}$ to $1.4 \times 10^{6}$; with $\epsilon = 0.01$, the required magic-state error rates range from $p_{out} \approx 6 \times 10^{-7}$ down to $7 \times 10^{-9}$, and are derived per benchmark in Table~\ref{tab:msd-requirements} of Appendix~\ref{app:resource-estimation}.
At the physical error rate of $p = 10^{-4}$, a widely studied single-stage distillation protocol called $15\text{-to-}1$ that takes 15 noisy T states and produces one magic state is sufficient for the benchmarks considered in this work, as a single-stage protocol already covers $n_T$ up to $\sim 10^{9}$ $T$ gates ~\cite{MSDNotCostly}. 
Substantially larger algorithms or higher physical error rates require either a concatenated distillation protocol such as a two-stage $15\text{-to-}1$ or $(15\text{-to-}1)\times(8\text{-to-}CCZ)$ ~\cite{MSDNotCostly} or a cultivation protocol such as fold-transversal cultivation~\cite{Sahay2025FoldTransversal}. 
While cultivation offers a significant reduction in space-time volume, its practicality may vary from platform to platform. 
For instance, the lowest-overhead cultivation variants require non-local or reconfigurable connectivity~\cite{Sahay2025FoldTransversal, HighRateCult}. 
Furthermore, for platforms like photonics, schemes to perform cultivation with hardware-native operations (e.g., photonic fusions~\cite{bartolucci2023fusion}) have yet to be developed, to the best of our knowledge. 
Therefore, we choose to introduce non-Clifford resources through zero-level distillation ~\cite{ZeroLevelDistillation} when the error rate is high ($p = 10^{-3}$), followed by a $15\text{-to-}1$ distillation. 
Table~\ref{tab:15-to-1-comparison} in Appendix~\ref{subapp:15-to-1-distillation} compiles the output error probabilities $p_{out}$ of the protocol variants used in this work at physical error rates of $p = 10^{-3}$ and $p = 10^{-4}$.
We elaborate more on the $15\text{-to-}1, $ circuit construction in Appendix~\ref{subapp:15-to-1-distillation}.
%


\section{Compilation architectures}
\label{sec:compilation-architectures}

\begin{figure*}
\centering
\includegraphics[width=1\linewidth]{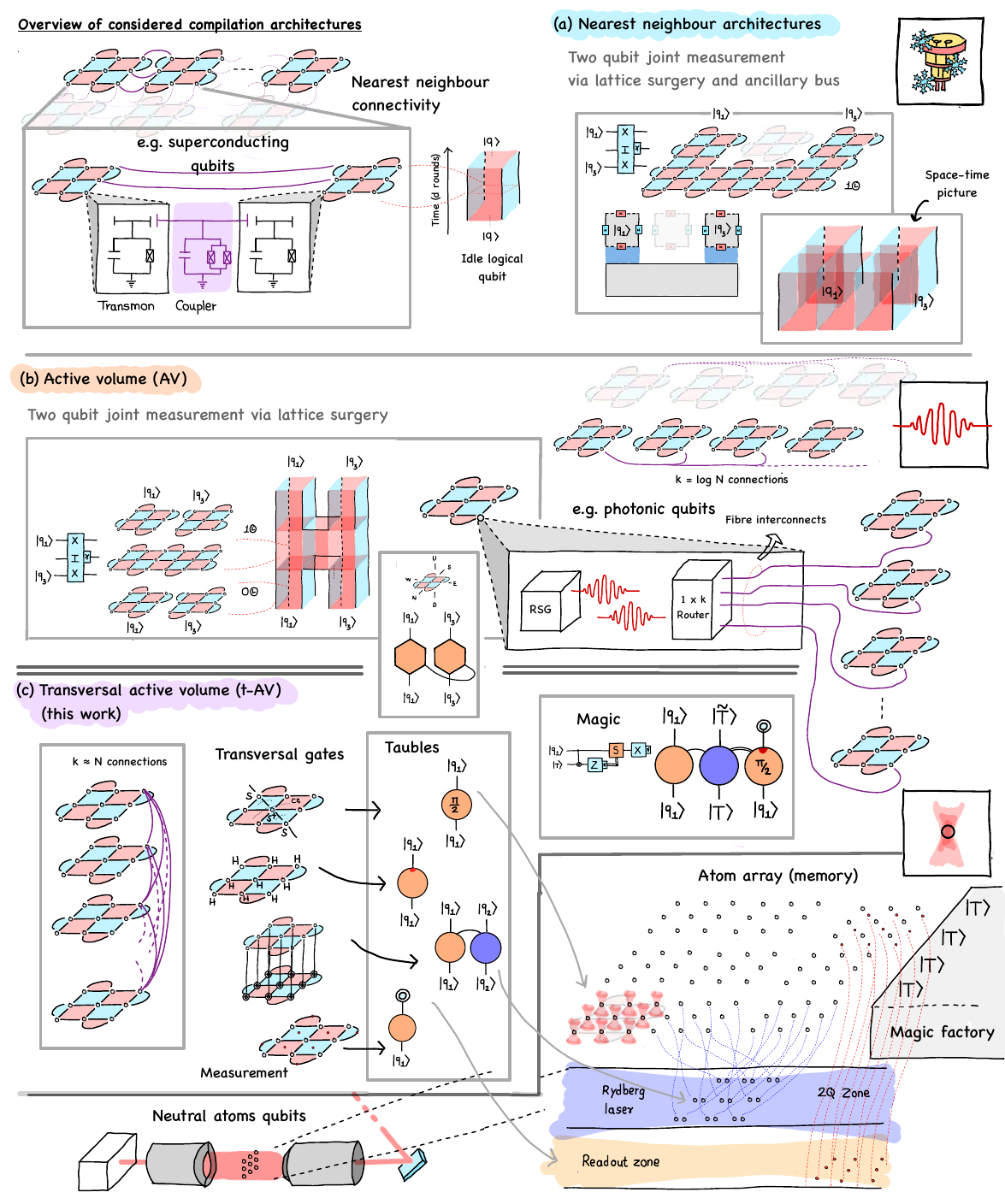}
\caption{Overview of compilation architectures. A logical qubit is a rotated surface-code patch (red/blue $=$ $X$/$Z$ boundaries).
(a) Superconducting platform with nearest-neighbor connectivity. (left)
Surface code patches sit on a static 2D grid with nearest-neighbour links (inset: transmon qubits and tunable couplers; the space-time diagram shows an idle logical
qubit as a $(d_X,d_Z,d_m)$ cube). (right) A joint PPM is routed through an ancillary bus via lattice surgery \cite{PBCGameSurface}, with a corresponding space-time picture.
(b) Photonics platform with limited non-local connectivity using active volume architecture \cite{AVLitinski}. Each patch carries $k=\log N$ long-range links realized by fiber interconnects
(inset: resource-state generator and $1\times k$ router).
The same PPM is performed by a direct merge/split between $q_1$ and $q_3$ and is represented as a logical network.
(c) Neutral atoms or photonic platforms with (effectively) all-to-all connectivity using the t-AV architecture (this work). With $k\approx N$ connectivity, t-Cliffords and measurements are compiled into logical blocks inspired
\emph{taubles}. See Sec.~\ref{subsec:transversal-active-volume-architecture} for more details.}
\label{fig:monsterfigure} 
\end{figure*}

We now briefly discuss the existing surface code-based compilation architectures built upon PBC in Sec.~\ref{subsec:quantum-computer-with-local-connectivity} and ~\ref{subsec:quantum-computer-with-limited-non-local-connectivity} and introduce our transversal active volume (t-AV) as a compilation architecture for platforms supporting t-Cliffords and effectively all-to-all connectivity in Sec. ~\ref{subsec:transversal-active-volume-architecture}.

\subsection{Platforms with local connectivity}
\label{subsec:quantum-computer-with-local-connectivity}
We consider the standard scenario of a fault-tolerant model based on the surface code with logical patches arranged on a static 2D layout with nearest-neighbor connectivity, as shown in Fig.~\ref{fig:monsterfigure}(a).
Here, we focus on the superconducting platform with physical qubits encoded in transmons and interconnected via physical couplers (limited to performing two-qubit gates between adjacent qubits), but this type of architecture is also relevant for other nearest-neighbor connectivity platforms such as semiconductor spin qubits. 
Architectures based on nearest-neighbor connectivity have been widely studied in the literature; see, for example, Refs.~\cite{PBCGameSurface, FowlerSurfFTQC, SurfCodeLattice}.
Because inter-patch operations are restricted to shared boundaries between adjacent patches, multi-qubit PPRs/PPMs between non-local qubits cannot be executed directly; they must instead be mediated by ancillary patches~\cite{SurfCodeLattice}.
Figure \ref{fig:monsterfigure}(a) shows the tile and space-time picture of a typical logical operation that is supported by platforms with local connectivity using the surface code: the execution of an $X_1 \otimes I_2 \otimes X_3$ PPM via lattice surgery using an ancillary bus.
The measurement is performed by merging and splitting the respective boundaries and takes $1 \clk$.
Such a configuration partitions the patch layout into three regions: a \textit{data block} hosting the logical qubits needed to encode the data, \textit{routing} ancilla patches to mediate logical operations such as PPMs, and a \textit{distillation factory} (not shown) that supplies high-fidelity $\ket{T}$ states.
These architectures have a huge space-time cost and idle qubit volume due to the use of routing ancillas.
Following Ref.~\cite{PBCGameSurface}, we consider two representative designs for this type of system. First, the \textit{baseline architecture} in which the data block contains surface code tiles that encode one logical qubit using two tiles, an ancillary bus of the length of the data block, as well as sufficient magic state factories to support one PPR execution per $1 \clk$. Second, the \textit{compact architecture}, in which the footprint of both the data block and the ancillary bus is minimized at the cost of a slower consumption rate of magic states.
These span the two extremes of the space-time trade-off available to a local-connectivity architecture.
The PBC architectures considered here and their space-time costs are discussed in greater detail in Appendix~\ref{subapp:baseline-architecture} and \ref{subapp:compact-architecture}.

\subsection{Platforms with limited non-local connectivity}
\label{subsec:quantum-computer-with-limited-non-local-connectivity}

The baseline architecture has a significant space overhead and idle volume due to locality restrictions, whereas the compact architecture has a lower space footprint at the expense of longer runtime and increased idle volume from slow consumption of magic states.
Platforms with limited non-local connectivity architectures allow us to use the idle volume to parallelize logical operations.
This section briefly describes one such widely studied architecture, the active volume (AV) architecture~\cite{AVLitinski}, that utilizes non-local connectivity to reduce the computational space-time volume.
As shown in Fig.~\ref{fig:monsterfigure}(b), the AV architecture can be supported, for instance, by a static 2D layout of photonic qubits with non-local connectivity. 
In this modality, qubits are encoded in the degree of freedom, e.g., polarization or path, of photons traveling in waveguides and optical fibers. 
We consider the computation to be performed in a fusion-based approach, that is, generating initial entangled states of photonic qubits through resource state generators (RSGs), and consuming them through two-qubit parity measurements --- linear optical fusions~\cite{bartolucci2023fusion}. 
In the fusion-based approach, optical switches and fiber interconnects are used to enable logical connectivity between different surface code patches, which can be implemented by fusing physical photonic qubits from RSGs associated with a logical patch with those of another patch~\cite{LogicalBlocks, AVLitinski, bombin2021interleavingmodulararchitecturesfaulttolerant}. 
A $k = \log N$ logical connectivity corresponds to interconnects that can route photons from the RSGs of that patch to $\log N$ other locations, i.e., physical optical routers with $1 \times \mathcal{O}(\log N)$ connectivity. 
The same $X_1 \otimes I_2 \otimes X_3$ PPM as discussed before can now be executed via a direct lattice merge and split between qubits $q_1$ and $q_3$ without the need for an ancillary bus, requiring one logical cycle $1 \; \clk$.
To better represent AV architecture logical operations, it is useful to consider the $(d_X, d_Z, d_m)$ space-time picture of a logical qubit as described in Sec. \ref{sec:platform-aware-compilation-and-metrics-of-interest}, and construct so-called logical networks using \textit{oriented}-ZX diagrams (see also Ref~\cite{AVLitinski} for more details) to represent corresponding logical operations. 
We show an example of the $X_1 \otimes I_2 \otimes X_3$ PPM as a logical network in Fig.~\ref{fig:monsterfigure}(b), and we discuss its construction and subsequent compilation in more detail in Appendix~\ref{subapp:active-volume-architecture}. 
The space-time volume in AV architecture is calculated by considering only the active part of the computation.

\subsection{Transversal active volume architecture}
\label{subsec:transversal-active-volume-architecture}
The AV architecture significantly reduces the idle space-time volume used by locality-constrained architectures in lattice-surgery-based computation.
However, the logical operations still scale with $d$, incurring extra error correction as well as space overhead.
As shown in Fig.~\ref{fig:monsterfigure}(c), if the platform instead supports effectively all-to-all connectivity and t-Cliffords (see Sec. \ref{subsec:t-clifford-t-computation}), as in, for instance, reconfigurable neutral-atom arrays where any pair of encoded surface code patches can be brought together into physical proximity via atom shuttling, or all-to-all connectivity between logical patches ($k = N$), for e.g., photonic qubit platforms where qubits are connected via fiber interconnects, a different compilation paradigm becomes available. 
The neutral-atoms platform considered here uses single atoms trapped in optical potentials that generally have programmable positions ~\cite{TweezerArray}, and which uses Rydberg-mediated entangling gates and single-site addressing ~\cite{FastQuantumGatesNeutralAtoms}.
The atoms can be dynamically assigned as data or ancilla qubits, where the ancilla qubits can be measured and moved back into memory for reuse. 
Since different operations are performed in physically distinct zones, this introduces significant time overhead on the order of $100~\mu$s for intermediate-scale systems ~\cite{Bluvstein2024LogicalProcessor}. 

On all-to-all platforms, instead of rewriting the circuit in terms of serial PPRs, one can directly execute the Clifford + $T$ circuit.
However, we cannot get any meaningful information about the space-time cost of each operation solely from the logical circuit, and therefore, we recompile it into an instruction set that can also incorporate the space-time volume of operations supported, such as t-Clifford, as well as distillation and injection overheads. 
Similar to the AV architecture, the space-time volume in the t-AV architecture is the volume of active logical operations, with the operations performed transversely.

In the following subsections, we outline the compilation of logical operations into ZX representation, describe their execution and parallelization, and finally discuss the reaction time overhead in our t-AV architecture. 

\subsubsection{ZX-diagram representation of transversal logical operations in t-AV architecture}
The $\mathrm{ZX}$-calculus offers a natural route to reason about quantum computations and circuits as graphs \cite{Coecke_2011}.
We briefly describe our use of $\mathrm{ZX}$-calculus and refer to Refs.~\cite{vandeWetering2020, Duncan2020, faulttoleranceconstruction, UnifyingFlavoursZX} for a more comprehensive introduction.
In ZX-calculus, quantum operations are represented as diagrams built from basic blocks called ``spiders" connected by wires. 
A Z(X)-spider with one input and one output wire represents a logical unitary transformation around the Z (X) axis, given by $\ket{0}\bra{0} + e^{i\alpha}\ket{1}\bra{1}$ ($\ket{+}\bra{+} + e^{i\alpha}\ket{-}\bra{-}$), where $\alpha$ is a phase. 
One can use these unitary transformations with discrete phases to represent the space-time picture of fault-tolerant logical operations. 
We express each transversal logical operation in its $\mathrm{ZX}$ representation and calculate the associated space-time volume as the number of logical qubits involved in the operation multiplied by the number of code cycles required to execute the operation, with units $ \boxdot \times \clk_c$.
In our architecture, the space-time cost to implement a given algorithm is hence the sum of the space-time volume of all the transversal operations taking part in the computation.
Below, we provide details on the STV calculation of transversal logical blocks. 

We model each active component of the recompiled circuit as a transversal logical block, which we refer to as a \textit{tauble} (transversal bauble), because, frankly, they look just like the little Christmas tree ornaments. 
Each tauble has an input and an output port and represents a transversal fault-tolerant logical operation. 
Fig.~\ref{fig:monsterfigure}(c) shows transversal Clifford gates and their representation as taubles with orange (blue) nodes representing the computational (Hadamard-transformed) basis. 
We use the same ZX color convention as in Ref.~\cite{AVLitinski}.
If a tauble carries no phase ($\alpha = 0$), it simply represents an idle transversal logical block/identity gate. 
The fold-transversal phase gate and its corresponding logical block are represented as a single-input and single-output orange tauble with a phase of $\alpha =\pi/2$, with space-time volume $1 \boxdot \times \clk_c$.
Similarly, the logical block corresponding to a Hadamard gate is represented as a single-input and single-output tauble (either blue or orange, depending on the logical operation), with the output port being Hadamarded (represented as a red dot), with space-time volume of $1 \boxdot \times \clk_c$.
Depending on the operation, a \textit{connect} port can be added between a pair of taubles, for example, when representing entangling gates. 
Using the ZX diagram of a CNOT gate, one can straightforwardly derive the taubles representation shown in Fig.~\ref{fig:monsterfigure}(c), where two taubles are connected to each other via a connect port in any direction. 
The connect port represents physical CNOTs acting on respective taubles. 
The resulting tauble pair has the space-time volume $2 \boxdot \times \clk_c$.
A $T$ gate is performed by first distilling and then injecting a $T$ state via an injection gadget consisting of a CNOT followed by an $S$ correction conditioned on a $Z$-measurement. 
The correction is applied with a $50\%$ probability, resulting in a space-time volume for $T$ injection of $3.5 \boxdot \times \clk_c$.
To quantify the total space-time volume of a computation, we also need to take the space-time volume of distillation or cultivation schemes into account. 
We derive these volumes in Sec. \ref{subsec:t-av-of-magic-factories-and-platform-specific-msd-layout-optimization}, followed by descriptions of platform-specific optimizations for the distillation protocols. 

\subsubsection{Execution of taubles}
To perform resource estimates on our t-AV architecture, we build a compiler that reformulates a given Clifford + $T$ logical circuit into a network of taubles. 
First, each circuit component is converted into taubles with up, down, and connect ports together with a phase, which together indicate the type of the transversal gate being executed. Each tauble is then assigned a unique serial number for scheduling the operations.
Taubles that act on a distinct set of qubits can be executed in parallel. 
Pairs of taubles acting on the same qubit cannot simply be parallelized without additional qubit overhead. 
Parallelizing such taubles can be achieved through the use of \textit{bridge qubits}, allowing the execution of two logical operations acting on the same logical qubit in parallel, in a given code cycle. 
Bridge qubits have been used in AV architecture compilers, as well as in compilers based on transversal gates~\cite{AVLitinski, ArchitectNeutralAtoms}, and we now describe how bridge qubits can be used in the t-AV architecture to parallelize the network of taubles.  
Two operations acting on the same data qubit can be parallelized by preparing a fresh Bell pair, with one half of the Bell pair (which we call a bridge) used to teleport the data qubit to the other half within the same cycle via Bell measurements. 
\begin{figure}[t]
\centering
\includegraphics[width=1\columnwidth]{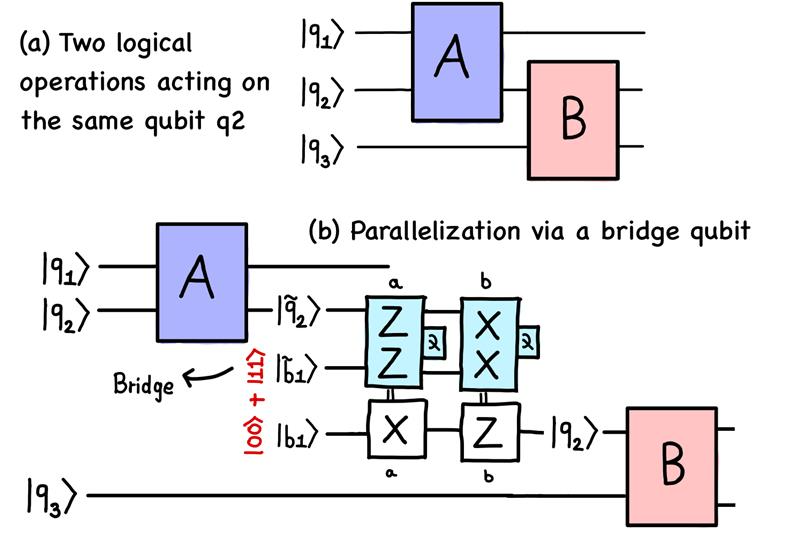}
\caption{Parallelizing two operations via Bell-state teleportation and bridge qubits}
\label{fig:bridge} 
\end{figure}
This procedure is depicted in Fig.~\ref{fig:bridge}(a).
Logical operations A and B correspond to a network of taubles acting on the same qubit $\ket{q_2}$.
To execute A and B within a single cycle (in parallel), a fresh Bell pair $\frac{1}{\sqrt{2}}(\ket{00} + \ket{11})$ is initialized on two bridge qubits, denoted by $\ket{b_1}$ and  $\tilde{\ket{b_1}}$, and performing a Bell measurement between $\tilde{\ket{q_2}}$ and $\tilde{\ket{b_1}}$ with a classically tracked Pauli correction teleports the qubit $\ket{q_2}$ to $\ket{b_1}$.
In our t-AV architecture, after scheduling a network of taubles, we search for taubles acting on the same qubits within a code cycle and allocate a bridge qubit when needed. 

\subsubsection{Reaction depth in t-AV architecture}
\label{subsec:reaction-depth-in-t-av-architecture}
As defined in Sec.~\ref{sec:platform-aware-compilation-and-metrics-of-interest}, a measurement whose basis depends on the outcome of a previous measurement is called a reactive measurement and has an associated reaction time $\tau_r$ and depth $R$.
These quantities set a runtime floor of $R \times \tau_r$ that no amount of spatial parallelism can reduce~\cite{AVLitinski, fowler2013timeoptimalquantumcomputation}.
In the t-AV architecture, the reactive event is the measurement within the T-injection gadget via transversal operations.
Each injection produces an $S$ correction on the target qubit.
When successive $T$ gates on the same qubit are separated only by Clifford operations whose action commutes with the pending $Z$-type correction, no new reaction layer is generated.
When an intervening Clifford rotates the correction into an anticommuting type, for example, a Hadamard that maps $Z \to X$ between two T injections, the next T-injection measurement basis anticommutes with the propagated correction, and a reaction layer is added.
The reaction depth per code cycle in the t-AV architecture is therefore bounded by the longest chain of T gates on any single qubit within that cycle in the worst case.

\subsection{t-AV of magic state factories and platform-specific MSD layout optimization}
\label{subsec:t-av-of-magic-factories-and-platform-specific-msd-layout-optimization}

The layout, as well as the execution time of distillation factories, is platform-specific and can be further optimized depending on the connectivity of the platform and the operations supported.
In this section, we outline a construction of magic state factories suitable for t-AV architecture and compute their space-time volume (STV). 
We discuss three variants. The first two variants are factories supporting transversal gates and all-to-all connectivity: namely, a transversal distillation (\textit{Trans-dist}) factory and a parity distillation (\textit{Parity-dist}) factory. 
Trans-dist factory relies on the original 15-to-1 CNOT construction (Fig.~\ref{fig:t-msd}), while Parity-dist relies on the reformulated PPM construction of 15-to-1 distillation (Fig.~\ref{fig:15to1pbc}), but with each physical operation performed transversely. 
The third variant that we consider is existing lattice surgery-based (\textit{LS-Dist}) factories for limited non-local connectivity in the distillation-specific zone.
In the last section, we describe a space-time efficient zero-level distillation that can be used to quadratically improve the magic state error rate when the physical error rate is high. 
We combine this zero-level distillation with different variants considered in this work and present it as an alternative to two-stage distillation when a benchmark requires an error rate that one-stage distillation cannot provide, which significantly reduces the space-time overhead. 

\subsubsection{Transversal $15\text{-to-}1$ distillation (Trans-dist)}
\label{subsubsec:transversal-15-to-1-distillation}
For platforms supporting transversal gate operations and all-to-all connectivity, distillation can be performed in constant time~\cite{ConstantTimeDistillation}.
Here we consider the CNOT construction of the $15\text{-to-}1$ MSD protocol with $[[15, 1, 3]]$ Reed-Muller code, as described in Fig.~\ref{fig:t-msd}, and compute its STV.
This $15\text{-to-}1$ circuit that can be executed in $9 \clk_c$ as follows: 
In code cycle $1$, all 15 MSD patches plus the magic state patch are initialized in their respective states (STV: $16 \boxdot \times \clk_c$).
Transversal CNOTs that form the encoding circuit take the next four code cycles ($2$--$5$) with STV of $50 \boxdot \times \clk_c$.
These CNOT circuits generate 
A Bell state of the 0th surface code patch entangled with the 15-qubit Reed-Muller code.
In code cycles $6$ to $8$, 15 noisy magic states (we assume that the noisy magic states can be prepared offline and ignore the overhead of initialization) are injected via gate teleportation (STV: $15\times3.5 \boxdot \times \clk_c$), with total STV of $52.5 \boxdot \times \clk_c$.
Code cycle $9$ performs $X$ measurements on the 15 data qubits,
in order to measure all four weight-8 $X$ stabilizer checks of the Reed Muller code, if all outcomes of the measurement are $+1$, a magic state is produced on the 0-th qubit, otherwise, the state is discarded and the protocol is repeated again (STV: $15 \boxdot \times \clk_c$), giving us a total STV of $133.5 \boxdot \times \clk_c$.
Based on these numbers, we can estimate the total STV for any given logical circuit, considering \textit{Trans-dist} as a distillation factory in a t-AV architecture.
Since each tauble corresponds to a
surface-code patch of $2d^2$ physical qubits held for one code cycle, a logical STV is converted into a physical space-time volume by multiplying by $2d^2$; for the 15-to-1 protocol, this gives $(133.5)\times 2d^2 = 267\,d^2$ physical STV per distilled $T$ state.

\textbf{Layout optimization:} On platforms with effectively all-to-all connectivity, the factory layout for the transversal 15-to-1 circuit can be further optimized.
Each factory has 15 logical qubits and 1 qubit where the magic state is produced, all encoded in surface code patches.
Consider $N_{fac}$ identical magic state distillation factories, each occupying
$16$ logical patches and producing one high-fidelity magic state every $9$ code cycles.
We start our magic state factories one code cycle apart, such that no two magic state factories are in the same factory state in any given code cycle for $N_{fac} \leq 9$.
At the 6th code cycle of each factory, we need to inject 15 noisy magic states via gate teleportation. 
We initialize an auxiliary 15-logical qubit register, a \emph{buffer bus}, into these noisy magic states, and use it for 3 consecutive code cycles (cycles $6$--$8$ of the factory schedule, see Fig.~\ref{fig:t-msd}).
The noisy states reside in the same 15 physical bus qubits throughout their 3-cycle window, so each factory requires the buffer bus for 3 \emph{consecutive} cycles, and they are free to be used after 3 code cycles for another factory. 
One bus can serve up to $\lfloor 9/3 \rfloor = 3$ factories if we start them one code cycle apart.
The minimum number of shared buffer buses required to avoid stalling in the production of magic states is $B_{\min}(N_{fac}) \;=\; \left\lceil \frac{N_{fac}}{3} \right\rceil$.
Each bus contributes $15$ logical qubits, so the buffer footprint is
$15\,B_{\min}(N_{fac})$ qubits, compared with $15N_{fac}$ qubits in the naive (non-shared) design.
For example, $N_{fac} = 9$ delivers one magic state per code cycle using $B_{\min} = 3$ buffer buses ($45$ qubits) instead of $9$ ($135$ qubits), a $66.7\%$ reduction in logical qubit overhead.
Therefore, for $N_{fac}$ magic state factories, with $B_{\min}(N_{fac})$ buffer buses, STV of $N_{fac}$ factories will be $(88.5 \times N_{fac} + 45 \times B_{\min}(N_{fac}))\times 2d^2$.

\subsubsection{Distillation with parity ancilla (parity-dist)}
\label{subsubsec:distillation-with-parity-ancilla} 
For architectures supporting transversal CNOTs, with effectively all-to-all connectivity, we use the 15-to-1 circuit construction derived from triorthogonal matrices (as shown in Fig.~\ref{fig:15to1pbc}), which distills one magic state by performing $15$ faulty $\pi/8$ rotations.  
However, instead of applying PPRs through lattice surgery, we initialize a parity ancilla and perform PPRs through transversal CNOTs.
Such a factory requires 7 logical qubits (5 factory qubits, 1 ancillary qubit, and 1 qubit for T state injection), and $k$ CNOTs per PPR, where $k$ is the Pauli-$Z$ weight of the PPR.
For example, for the following PPR: $(I\otimes Z \otimes Z \otimes Z \otimes I)_{\pi/8}$, one requires $5$ CNOTs and hence $5 \clk_c$ for the execution.
For any distillation protocol derived from triorthogonal matrices, if the protocol contains $m$ PPRs, the total number of CNOTs required is $\sum_{i = 1}^m k_i$, where $k_i$ is the number of Pauli $Z$s in the $i$-th PPR. 
In the 15-to-1 protocol, with the first 4 PPRs absorbed as $T$-state preparation on qubits 1--4, we need $35$ sequential CNOTs, taking $35 \clk_c$.
Conventionally, a T injection would take $4 \clk_c$, with a STV of $3.5 \boxdot \times \clk_c$; with PPR execution, a noisy T state can be prepared in parallel while other CNOTs are being executed, hence we count only 3 code cycles for injection, and the STV remains the same, so for 11 rotations we need $22$ code cycles for injection with a STV of $11\times 3.5 \boxdot \times \clk_c$.
Since each CNOT has an STV of $2 \boxdot \times \clk_c$, the total STV of the Parity-Dist factory becomes  $108.5 \boxdot \times \clk_c$, taking $60 \clk_c$ (one for initial state preparation, one for final measurement).
The space-time volume for parity-dist is therefore $108.5\times2d^2 = 217d^2$ per $T$ state distillation.

\subsubsection{Lattice-surgery based distillation factories (LS-dist)}
\label{subsubsec:lattice-surgery-based-distillation-factories}
Lattice surgery-based factories are well-suited for limited non-local ($\log N$)/local connectivity (See Sec. \ref{sec:platform-aware-compilation-and-metrics-of-interest} for connectivity definitions). 
We also consider a scenario where magic state generation is performed with $\log N$ connections and lattice surgery, and the actual computation is performed transversely with effectively all-to-all connections between the surface code patches, to quantify how much improvement we gain solely from transversal operations as compared to lattice surgery-based AV architecture \cite{AVLitinski}. 
With a platform supporting $\log N$ non-local connections, the standard variant of the $15$-to-$1$ distillation protocol, $(15\text{-to-}1)_{d, d/2, d/2}$, of Fig.~\ref{fig:15to1pbc} can be laid out on 35 logical qubits as a network of $35$ half logical blocks of dimension $(d, d, d/2)$ (See Fig. 20 of Ref. \cite{AVLitinski} for detailed layout).
Such a logical network produces $1$ T state per half a logical cycle.
In this case, the physical space-time volume of the computation is simply $35 \times 2d^2 \times d/2 = 35d^3$ per $T$ state. 

\subsubsection{Concatenation of MSD protocols with Zero-level distillation}
\label{subsubsec:concatenation-of-msd-protocols-with-zero-level-distillation}
Whenever a benchmark requires a magic-state error rate that $15\text{-to-}1$ alone cannot provide, concatenated distillation (e.g., a two-level $15\text{-to-}1$ factory) becomes resource-expensive: the 15 input states of the second level are no longer free noisy states but each cost a full first-level run, so with each $15\text{-to-}1$ factory contributing an STV of $267d^2$, the two-level factory totals $16 \cdot 267d^2 = 4272d^2$ per output $T$ state.
To avoid this blow-up in the space-time volume of distillation, we use zero-level distillation (0-dist) to produce magic states of slightly higher fidelity, which are then fed to $15\text{-to-}1$. 
0-dist has a relatively lower space-time volume than a single 15-to-1 factory~\cite{ZeroLevelDistillation}.
Here, we first briefly describe 0-dist and propose its merging with different distillation variants considered for t-AV: namely, Trans-dist, LS-dist, and Parity-dist.

\textbf{Zero level distillation (0-dist)} 
performs the Hadamard test of the logical $\mathrm{H}$ gate on the seven-qubit Steane code to distill a magic state; the distilled state is either teleported to the rotated surface code, or the Steane code itself is fault-tolerantly converted into a rotated surface code, and the process is called code conversion\cite{ZeroLevelDistillation}.
Teleportation-based 0-dist requires 40 physical qubits that can sit on a 5$\times$5 surface-code patch and requires 5 code cycles to produce a magic state, with a discard rate of $30\%$ \cite{ZeroLevelDistillation}.
Code-conversion-based 0-dist requires only 15 physical qubits and produces a magic state per 8.4 code cycles with a similar discard rate \cite{ZeroLevelDistillation}.
We urge readers to look at Sec. III and IV of Ref. \cite{ZeroLevelDistillation} for details on derived numbers.
Therefore, in total, it takes 12 code cycles, and growing the code to the given distance of the data qubit patch (1 code cycle) introduces approximately 13 code cycles to produce a zero-level magic state. 
For an error rate of $p = 10^{-3}$, the 0-dist protocol produces a magic state with an error rate of $p = 10^{-4}$ \cite{ZeroLevelDistillation}.

\textbf{0-Dist + Trans-Dist:}
In 15-to-1 Trans-dist, a magic state is injected at the 6th code cycle using a 15-qubit buffer bus (See Sec. \ref{subsubsec:transversal-15-to-1-distillation}).
We perform 0-Dist directly on the buffer bus qubits via code conversion. 
0-Dist with code conversion takes 13 code cycles in total (taking into account the discard rate) to produce a zero-level magic state. 
This state is then fed to the 15-to-1 transversal protocol. 
It does not require additional space overhead, but an extra 13 code cycles are added per magic factory, due to the use of 0-dist.
Therefore the STV of 0-Dist becomes $(15)\times 13 = 195 \boxdot \times \clk_c$ and the total STV of the concatenated protocol becomes $(133.5+195) = 328.5 \boxdot \times \clk_c$, giving total space-time volume of $657d^2$.

\textbf{0-Dist + Parity-Dist:}
While concatenating 0-Dist with Parity-Dist, we do not absorb the first four rotations into a $T$ state preparation (See Fig. \ref{fig:15to1pbc}), as they need to be explicitly prepared through 0-Dist to reach higher fidelity. 
Similar to the Trans-Dist case, we use code conversion; here, 15 injections are performed sequentially; therefore, an extra $15 \times 13$ code cycles are added to the distillation factory.  
The STV thus becomes $(108.5 + 195) = 303.5 \boxdot \times \clk_c$, giving a total space-time volume of $607d^2$.

\textbf{0-Dist + LS-Dist:}
With LS-Dist, one can either concatenate 0-Dist via lattice surgery or directly grow the state on a surface code patch. 
With the first approach, concatenation of 0-dist with LS-Dist produces a magic state with an error rate of $5.7 \times 10^{-11}$ with an input error rate of $p = 10^{-3}$ \cite{MSDNotCostly}. 
We assume a factory layout given in Ref. \cite{Hirano_2024} with the following parameters:
We perform zero-level distillation in a $d_Z \times d_Z$ patch, requiring $d_Z$ code cycles, and we use 4 such blocks to produce one magic state per 7 logical cycles.
With standard 15\text{-to-}1$_{d_X, d_Z, d_m}$, we do not need ancillas in their layout for non-local connections.
We assume $d_X$ to be whatever code distance the data block uses, and assume $d_Z = 11$ as per their construction.

To directly grow the states on surface codes, via code conversion, as we mentioned before, it takes 13 code cycles to produce a zero-level magic state on 15 $d \times d$ surface code patches. 
Considering the standard logical block layout (Fig. 20 of Ref. \cite{AVLitinski}) of a $15\text{-to-}1_{d, d/2, d/2}$ factory, we inject 15 noisy magic states into 15 respective PPMs. 
In their current construction, the first four rotations are absorbed into their respective qubits and treated as noisy $T$ state initialization. 
However, when concatenating 0-dist with LS-dist, instead of initialization, we perform zero-level distillation on them. 
Therefore, these 4 surface code patches need to have a full spatial distance $d \times d$ instead of $d \times d/2$ as per their original construction; all the other half blocks connected to them also need to have the full spatial distance (Fig. 20 of Ref. \cite{AVLitinski})). 
As there are 8 such blocks, we need 8 more half blocks to perform 0-dist with LS-dist, requiring 43 half blocks in total for half a logical cycle, with total time to execute it as $d/2 + 13$ code cycles.
Therefore, the total space-time volume of 0-dist + LS-dist is $390d^2 + 43d^3$.

\subsubsection{Fold-transversal cultivation (cult)}
\label{subsubsec:fold-transversal-cultivation}
Due to the high space-time volume required by distillation factories, non-Clifford computation is expensive.
A significant part of the volume can be reduced by cultivating magic states instead of distilling them, which have substantially lower space-time overhead.
For t-AV, we consider performing cultivation with operations that enable reconfigurable connectivity, which is suitable, e.g for implementations on neutral atom arrays.
In particular, we use fold-transversal surface-code cultivation, which is suitable for architectures with reconfigurable connectivity and produces a magic state by measuring a transversal logical Clifford operator on an initial small-distance code and then rapidly growing to a larger-distance code ~\cite{Sahay2025FoldTransversal}.
We then provide space-time volume for different error rates, and details on how the numbers are derived are given in the Appendix~\ref{subapp:fold-transversal-surface-code-cultivation}.
In this scheme, high-fidelity $\ket{T}$ states are prepared on the rotated surface code by measuring the fold-transversal $H_{XY}(=(X+Y)/\sqrt{2})$ on the mid-circuit unrotated surface code, and unitary as well as stabilizer measurement techniques are used to rotate and grow the code.
We compute the space-time volume for two different error rates.
With $p = 10^{-3}$, the error rate reached by fold-transversal cultivation is $10^{-8}$ with 5 expected attempts and 12 code cycles per attempt, giving us 60 code cycles per state (See Appendix~\ref{subapp:fold-transversal-surface-code-cultivation}). 
Note that this is an upper bound, and in practice, time could be much lower, because post-selection can succeed in fewer than the maximum code cycles required, leading to lower time per cultivated T state\cite{Sahay2025FoldTransversal}. 
The space cost is $562 + d_{\mathrm{eff}}^2$ physical qubits, where $d_{\mathrm{eff}} = \min(d, 11)$ is the cultivation patch distance (saturating at $d_{\mathrm{eff}} = 11$, i.e. $683$ qubits, for $d \geq 11$).
Therefore the spacetime cost is $(562 + d_{\mathrm{eff}}^2)\cdot 60$ per $T$ state, saturating at $683 \cdot 60 \approx 4.09 \times 10^4$.
For an error rate of $p = 10^{-4}$, we obtain 18 code cycles per magic state with the same physical cost. 

\begin{table*}[tbp]
\centering
\caption{Summary of important specifications and assumptions for the different hardware platforms considered here.}
\begin{tabular}{l c c c}
\hline\hline
Specifications & Superconducting & Photonics & Neutral-atoms\\
\hline
Connectivity       & Nearest-Neighbor & $\log N$/all-to-all & All-to-all \\
Compatible architecture $\;$    & Baseline, Compact \cite{PBCGameSurface}  & AV\cite{AVLitinski}/t-AV (this work) & t-AV (this work) \\
Code cycle time ($\tau_c$)   & $1\mu$s \cite{PBCGameSurface}  & $1\mu$s\cite{AVLitinski} & $10$ ms \cite{NeutralAtomCodeCycle} \\
Noise model  & Circuit-level(\ref{circ-noise-model}) $\;$ & Circuit-level(\ref{circ-noise-model}) $\;$ & Erasure-conversion (\ref{erasure-noise-model}) \\
\hline
\hline
\end{tabular}
\label{tab:platforms}
\end{table*}

\section{Results}
\label{sec:results}

\begin{figure*}[!t]
\includegraphics[width=\linewidth]{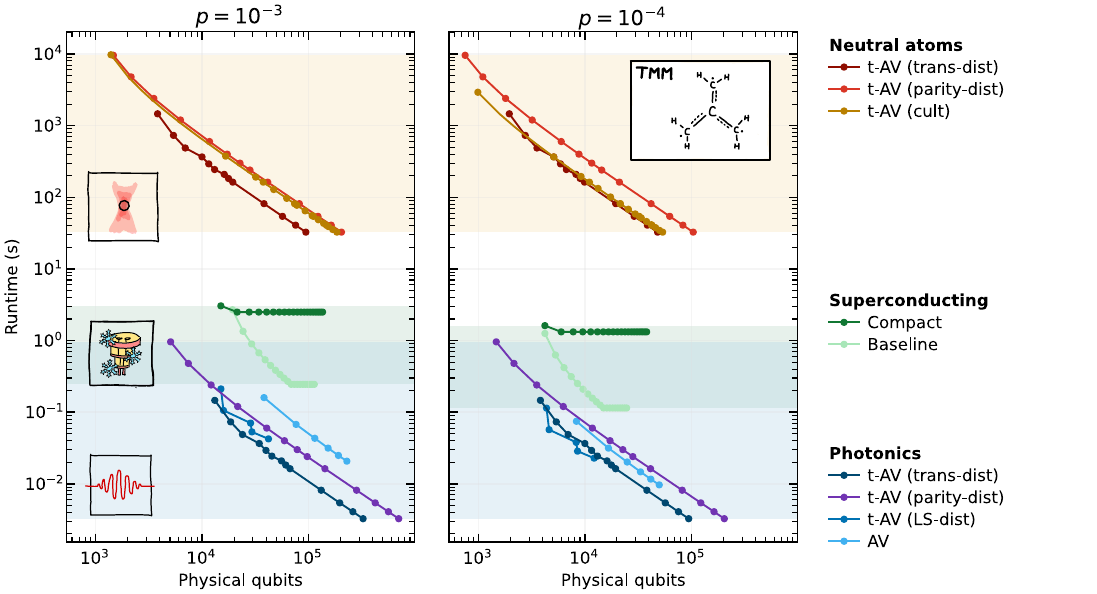}
\caption{Runtime vs. physical-qubit count for the Stat-QPE-gap TMM use benchmark, with $r_{max}=20$ and $p=10^{-3}$ (left) and $p=10^{-4}$ (right).
Each curve represents runtime-resource tradeoffs within one architecture-hardware configuration and is generated by sweeping the scheduling parameter specific to the architecture: workspace capacity for AV, T-per-cycle for t-AV/cult-t-AV, and factory count for Baseline/Compact. We specify the considered architecture-hardware configurations in the legend to the right of the plots. For the t-AV architecture, we also specify different magic state factories that are compatible with those architecture-hardware configurations. For readability, the plot is divided into three shaded areas by hardware platform: neutral atoms (sand), superconducting (green), and photonics (blue).}
\label{fig:runtime-vs-count-tmm}
\end{figure*}

So far, we have discussed fault-tolerant architectures that can be tuned to the capabilities of different hardware, focusing particularly on superconducting, photonics, and neutral atoms platforms (see Fig.~\ref{fig:monsterfigure}); however, other platforms can also be considered for this compilation framework, satisfying one of the connectivity types.
We now present our compilation pipeline applied to specific problems, which makes it possible to see cross-platform trade-offs and to compare the performance of our t-AV architecture against the other compilation architectures considered in this work.
We first discuss some important hardware assumptions made for our resource estimates and summarize them in Table~\ref{tab:platforms}. 
We assume a code cycle time of $1\,\mu\text{s}$ for both the superconducting and the photonics platforms \cite{PBCGameSurface, AVLitinski}. 
For the neutral atoms platform, we assume a code cycle time of $10\,\text{ms}$ \cite{NeutralAtomCodeCycle, HighFidelityGates, UniversalQuantumOperations}. 
We note that operation times are continually evolving and depend heavily on the choice of qubits as well as on trade-offs between speed and fidelity.
Recent developments in neutral-atom readout techniques have demonstrated readout timescales of $10$--$100,\mu$s \cite{MSScaleAtoms, UltrafastReadoutAtoms}, potentially reducing the code cycle time by an additional $2$ orders of magnitude.
However, in this work, we use a $10,\text{ms}$ code-cycle time to enable a direct comparison with the resource estimates reported in Ref.~\cite{khan2026architectingearlyfaulttolerant} for the neutral-atom architecture.

Noise model assumptions for each platform are discussed in Appendix~\ref{subapp:noise-model-and-distance-calculation}. 
Code distances per benchmark are calculated in Appendix~\ref{subapp:noise-model-and-distance-calculation} and summarized in Table~\ref{tab:min-distance}. 
We pair the distillation factories outlined in Sec.~\ref{subsec:t-av-of-magic-factories-and-platform-specific-msd-layout-optimization} with a compatible hardware platform in t-AV: lattice surgery (LS-Dist) is suitable for photonic platforms with limited connectivity, cultivation (Cult) is suitable for neutral atoms, and parity distillation (Parity-Dist) and transversal distillation (Trans-Dist) are suitable for both photonics and atoms.
Our assumptions related to magic state factories for the baseline, compact, and AV architectures are outlined in Appendix \ref{app:architectures}.

\begin{figure*}[!t]
\includegraphics[width=\linewidth]{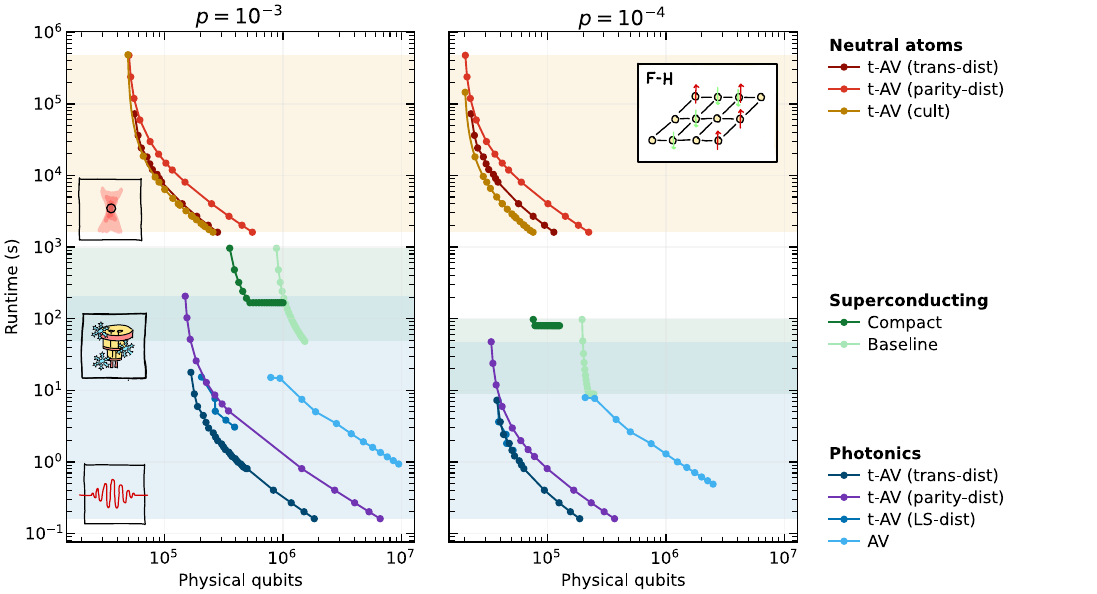}
\caption{Runtime vs. physical-qubit count for the Fermi-Hubbard benchmark at $p=10^{-3}$ (left) and $p=10^{-4}$ (right). See the caption of Fig.~\ref{fig:runtime-vs-count-tmm} for further details. Photonic t-AV reaches the lowest runtimes, while neutral-atom t-AV (cult) achieves the lowest physical qubit count, at the cost of a significantly slower code cycle.}
\label{fig:runtime-vs-count-fh}
\end{figure*}

We obtain end-to-end resource estimates for two concrete quantum simulation problems to compare the different architecture-hardware configurations. 
First, we consider the calculation of eigenenergies of a small chemical system for three cases of Trotterized phase estimation: 1) standard quantum phase estimation (QPE-Abs) to calculate the ground state energy (absolute energy calculation), 2) statistical phase estimation (Stat-QPE) to calculate the ground state energy, and 3) statistical phase estimation to calculate the energy gap between the ground and the first excited state (Stat-QPE-Gap).
Second, we compile a single 4th-order Trotter step of a $10\times 10$ square lattice Hubbard model, an application recently considered in Ref~\cite{khan2026architectingearlyfaulttolerant}. 
Specifically, we implement the dynamics of a symmetry-shifted Fermi-Hubbard model according to Refs.~\cite{Campbell_2021, AndreasTileTrotter:D}, and choose $U /\tau = 8$ for the system parameters. 
We describe these two problems and the construction of their respective logical Clifford + $T$ circuits in Appendix~\ref{app:quantum-simulation-problems}.
The Stat-QPE-Gap and the Fermi-Hubbard results are presented in Fig~\ref{fig:runtime-vs-count-tmm} and~\ref{fig:runtime-vs-count-fh} in the main text, while the Stat-QPE and QPE-Abs results are presented in Figs~\ref{fig:runtime-vs-count-stat-qpe} and~\ref{fig:runtime-vs-count-qpe-abs} in Appendix~\ref{subsubapp:stat-qpe-resource-estimates}. 
In these figures, we show estimates of the number of physical qubits required, including system qubits, distillation factories, and bridge qubits used for parallelization of logical operations, against time-to-solution (runtime) per benchmark.
In these resource estimate figures, each compilation architecture occupies a distinct region of the runtime--qubit plane: neutral atoms (sand), superconducting (green), and photonics (blue). 
In all of our resource estimates, the neutral atom computations have a slower runtime than the other platforms due to slower code cycle times. 
For each architecture, platform, and magic state factory pairing, for example, in the case of neutral atoms with t-AV architecture using a Trans-dist magic state factory. 
We provide several resource estimates (along each curve with fixed platform, architecture, and magic state factory) that are either in the magic-limited or the compute-limited zone. 
The magic-limited zone is where the production of magic states is the bottleneck. 
We can then reduce the runtime by increasing the number of factories (and the number of physical qubits) until the product of the number of factories and 1/production-time is 1.
After this point, the computation is not magic limited, and becomes compute limited.
In this zone, one can parallelize operations via bridge qubits (see Fig. \ref{fig:bridge}) such that the number of T gates per code cycle matches the number of magic states produced.
Below, we present our TMM Stat-QPE-Gap and Fermi-Hubbard results in more detail.

\textbf{TMM, Stat-QPE-Gap benchmark:} 
For this benchmark, the $T$-state error rate must be below $6.2 \times 10^{-7}$, and therefore, a single-level 15-to-1 distillation is sufficient at both $p = 10^{-3}$ and $p = 10^{-4}$.
For the longest circuit in the TMM Stat-QPE-Gap use-case with neutral atoms ($p = 10^{-3}$), both t-AV (parity-dist) and t-AV (cult) require $\sim10^3$ physical qubits and $10^4$s of runtime with magic limited computation (using the fewest number of physical qubits possible). 
On the other hand, the t-AV (trans-dist) curve requires less runtime ($10^3$s with $\sim10^4$ physical qubits) in the magic-limited zone, due to the distillation cycle being faster, but with larger qubit overhead compared to the other two variants. 
As we increase the number of factories, runtime evidently goes down.
All three variants follow similar trends since the computational cost is dominated by distillation factories rather than data qubits, and both factories have comparable space-time at this scale. 
The t-AV (cult) curve shifts to the left (towards fewer physical qubits) when the error rate is improved to $p = 10^{-4}$ due to the improved success probability per attempt, requiring less time per cultivation round.
We see a roughly linear improvement in runtime when increasing the number of physical qubits, and the computation does not saturate in runtime due to TMM being such a small system.
Photonic t-AV variants follow similar trends for t-AV with parity-dist and trans-dist, but with four to five orders of magnitude lower runtime due to faster code cycles.
The resource estimate curves are shifted slightly towards more physical qubits under the photonics noise model, which is a conventional circuit-level noise model, as opposed to the erasure-qubit conversion in atoms.
We also consider another variant, t-AV (LS-Dist), where distillation is performed via lattice surgery, which is useful when parts of a device have limited non-local connectivity. 
The AV architecture resource estimates are comparable to the t-AV architecture estimates for this example because the system is small and the major overhead comes from distillation factories.

The superconducting baseline and compact architectures fall between the neutral atoms and photonics platforms in runtime, and their resource estimates behave differently as we increase the number of physical qubits. 
For example, for the compact architecture, increasing the number of factories hardly improves the runtime. 
This is because even if factories are abundant, the runtime is limited by $T$-state consumption by the data blocks.
For the baseline architecture, since $T$ states are consumed one per logical cycle, the runtime decreases when adding factories, but once a sufficient number of factories is reached to keep up with the $T$-state consumption rate (one per logical cycle), the runtime hits a plateau. 
From these points, it is not possible to further parallelize the operations in such connectivity-constrained architectures.

\textbf{Fermi-Hubbard benchmark:} 
This benchmark problem requires roughly two orders of magnitude more qubits than the TMM benchmark, and at this scale, the dominant space overhead is no longer distillation but the number of physical qubits required to encode logical data qubits. 
The Fermi-Hubbard benchmark requires a T-state error rate below $1.2 \times 10^{-8}$. 
For t-AV with atoms, erasure conversion enables efficient post-selection, so that a physical error rate of $10^{-3}$ yields an injected T-state error of $10^{-4}$. 
A single round of 15-to-1 distillation, therefore, allows us to stay well under the budget. 
However, for photons, no such post-selection is available, meaning the injected error remains at $10^{-3}$. 
Therefore, 15-to-1 distillation alone cannot provide a sufficiently low error. 
Instead, we use the 0-dist + distillation variants of our magic state factories as explained in Sec.~\ref{subsubsec:concatenation-of-msd-protocols-with-zero-level-distillation}. 
For t-AV with neutral atoms, there is a clear separation between the parity and cultivation curves caused by this benchmark system being significantly larger, indicating that cultivation scales better with system size than parity distillation.
The t-AV (trans-dist) curve follows a similar trend.
In the photonics region, we see a clear distinction between the t-AV architectures and the AV architecture, where t-AV requires approximately an order of magnitude fewer physical qubits at $p=10^{-3}$, and almost two orders of magnitude fewer physical qubits at $p=10^{-4}$.

This significant reduction comes from fast transversal operations. 
AV, by contrast, relies on lattice surgery, which comes with significantly more overhead.
As in the TMM-QPE-Gap benchmark, the runtime of the superconducting baseline and compact architectures eventually hit a plateau when the computation becomes compute-limited. 

\section{Discussion}
\label{sec:discussion}
\subsection{Parallelization and Bridge-qubit demand}
\label{subsec:parallelization-and-bridge-qubit-demand}
\begin{figure}[tbp]
\centering
\includegraphics[width=0.9\columnwidth]{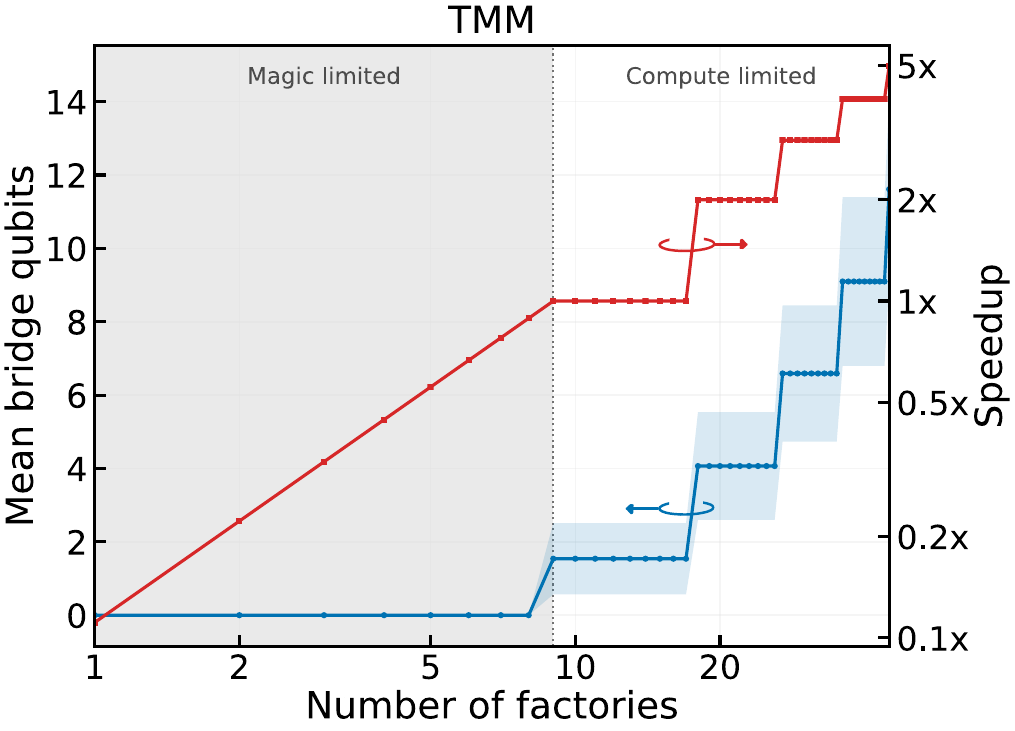}
\caption{Runtime-resource tradeoff in the t-AV architecture with varying number of $T$-state factories and bridge qubits of a single TMM-PPP model Trotter step. 
The blue curve (left axis) shows the mean bridge qubit demand per code cycle, with the shaded band spanning $±1$ standard deviation. 
The red curve (right axis) shows the speedup in logical cycle count when T factories and bridge qubits are added. 
The speedup is measured relative to the crossover point between the magic-limited and compute-limited zones (the $1\times$ mark). Adding T-gate factories reduces total runtime by $\sim 1/N_{fac}$, but sustaining these savings beyond the crossover point requires multiple $T$ injections on the same qubit per cycle. Then, the mean bridge qubit demand grows approximately linearly with $N_{fac}$, with $\approx 2.5$ bridge qubits per additional $T$ state supplied per cycle, since logical qubits must be teleported onto bridge qubits between consecutive operations.}
\label{fig:bell pairs t-av}
\end{figure}

Parallel execution reduces runtime, either when logical operations act on disjoint sets of qubits or when the architecture can supply auxiliary Bell pairs (see Fig. \ref{fig:bridge}) for routing logical information between simultaneously scheduled operations and performing Bell measurements within a single code cycle.
In t-AV architecture, execution of taubles complete within a single code cycle, making the runtime far more sensitive to magic state production rate.
For example, magic state production with a single Trans-dist factory takes $9 \clk_c$, so the next non-Clifford logical operation needs to wait for the factory before execution. 
Adding bridge qubits to the system only increases the idle volume of the computation, without enabling additional parallel execution. 
Therefore, we only introduce bridge qubits when the computation hits the compute-limited zone. 
Figure~\ref{fig:bell pairs t-av} shows the trade-off between magic state factories, bridge qubits, and runtime with a Trans-dist factory. 
We measure the speedup relative to the crossover point between the magic-limited zone and the compute-limited zone, which occurs when the number of factories is at $N_{fac} = 9$.
Starting from a single factory, the speedup increases linearly as the number of $T$-state factories increases until the computation becomes compute limited zone at $N_{fac} = 9$. 
At this point, the factories deliver one magic state per code cycle (magic state production rate is 1), and then the computation is limited by consumption of magic states (which takes $3 \clk_c$) rather than production.
In the compute-limited zone, multiple T gates can be scheduled in parallel, depending on the number of magic states available. If the T gates act on different qubits, they can be executed simultaneously. However, in the benchmark systems considered here, and typically in Trotter-based quantum simulation, arbitrary rotations are decomposed as sequences of T+Clifford gates, which makes it likely to have multiple consecutive T gates acting on the same qubit. 
Such operations cannot be parallelized without bridge qubits, so increasing the number of factories provides no further speedup.
However, this is solved by adding bridge qubits, giving up to a $5\times$ speed-up compared to the compute-limited zone crosspoint, at the cost of only a few bridge qubits per cycle. 
The TMM and Fermi-Hubbard t-AV benchmarks show similar speedup gains from adding bridge qubits, since their compiled circuits have comparable local gate structure. 
The bridge-qubit to speedup tradeoff is governed by local circuit structures rather than problem sizes.

\subsection{Reaction time estimates}
\label{subsec:reaction-time-estimates}
\begin{figure}
\centering
\includegraphics[width=\columnwidth]{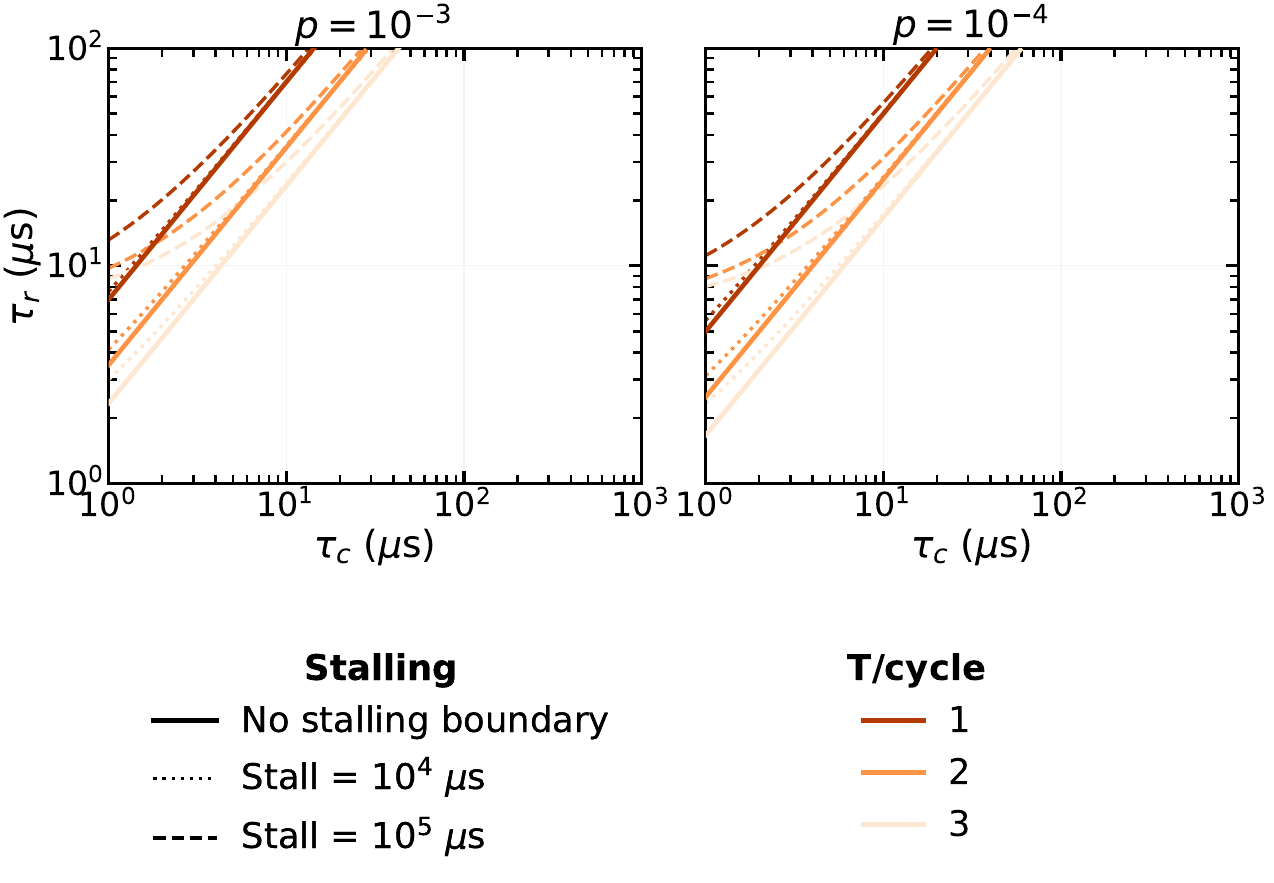}
\caption{Reaction-time stalling phase diagram for the TMM Stat-QPE-Gap benchmark in t-AV  architecture, shown for physical error rates $p=10^{-3}$ (left) and $p=10^{-4}$ (right). 
The solid line in each panel marks the no-stalling boundary $\tau_r\cdot k_{\max} = d\,\tau_c$ for three different T/cycle, and below these lines, the classical reactive processing completes within one logical cycle, meaning the computation is not reaction-limited. Dotted and dashed contours show iso-stall regions for total accumulated stalling time of $10^4\,\mu$s and $10^5\,\mu$s.}
\label{fig:reaction-time-tmm}
\end{figure}

The runtime of a fault-tolerant quantum computation can also be limited by the reaction time, instead of, for example, magic state production. 
Fig.~\ref{fig:reaction-time-tmm} shows stalling phase diagrams of the TMM Stat-QPE-Gap benchmark.
In each panel, the solid lines (indicated for three different T states per code cycle) show the no-stalling boundary, given by $\tau_r \cdot k_{\max} = d\,\tau_c \;,$
where $\tau_r$ is the reaction time, $k_{\max}$ is the worst-case per-cycle reaction depth, $d$ is the code distance and $\tau_c$ is the cycle time.
Because each $T$-injection correction is $Z$-type and can be deferred into the adaptively chosen basis of a later auxiliary $\ket{Y}$ measurement rather than applied in place~\cite{ArchitectNeutralAtoms}, corrections accumulate in the classically tracked frame and need only be resolved at the logical-cycle boundary. 
The classical processing of the $k_{\max}$ reactive layers arising within one logical cycle therefore has the full duration $d\,\tau_c$ of that cycle to complete, at the cost of one auxiliary patch per deferred correction.
The dotted and dashed contours mark iso-stall regions with \textit{total} stalling overheads of $10^4\,\mu$s and $10^5\,\mu$s, respectively.
The neutral atoms and photonic platforms have very different time scales for syndrome extraction ($\tau_c \sim 1\,\mu$s for photonics, $\tau_c \sim 10\,\mathrm{ms}$ for atoms). 
This results in several orders of magnitude more reaction-time budget for the neutral atoms platform at the same reaction depth.
Assuming a reaction time of $\tau_r =10 \mu s$ \cite{heavey2026scheduleractivevolumearchitecture}, photonic t-AV sits close to the no-stalling boundary at 1 T/cycle as shown in Fig.~\ref{fig:reaction-time-tmm}, and we therefore expect none or negligible overhead in runtime.
As the T state supply increases, stalling becomes prominent and eventually becomes a limiting factor for the photonics t-AV computation runtime.
For the neutral atoms platform, these computations lie well inside the no-stalling boundary, so reaction depth contributes no overhead even at higher T-state consumption.
Improving the physical error rate reduces the code distance, which tightens the requirement on the reaction time. Reaction-limited computation becomes more likely in this regime, and as seen in the right panel of Fig.~\ref{fig:reaction-time-tmm}, photonic t-AV incurs a total stall time of $\sim10^5 \mu$s even at 1T/cycle, while neutral-atom t-AV remains well inside the no-stalling zone.

\subsection{Space-time volume estimates}
\label{subsec:space-time-volume-estimates}
Space-time volume (STV) is a measure of the overall implementation cost and is calculated differently across platforms and architectures.
For the superconducting platform with baseline and compact architecture, STV is simply the product of the number of logical qubits and the total runtime in code cycles. In these architectures, routing ancillas are part of the active computation in each logical cycle and hence incur additional error correction overhead. 
On platforms with either limited non-local connectivity or effectively all-to-all connectivity, one exploits the non-local connections to reduce the routing overhead, and it is therefore sufficient to only consider the active part of the computation when calculating the total STV - the active volume for AV architecture or the transversal active volume for the t-AV architecture.
Table \ref{tab:stv-summary} summarizes the space-time volume for each architecture across the benchmarks considered in this paper. Neutral atoms t-AV achieves the lowest STV for all the benchmarks, followed by t-AV on photonics hardware.
\begin{table*}
\caption{\label{tab:stv-summary}Per-benchmark STV of each compilation architecture at two physical error rates, all in physical qubits $\times$ code cycles. The (t-AV) AV rows report the (transversal) active volume, and the Baseline and Compact rows report the minimum spacetime volume. 
Bold entries indicate the minimum STV in each column.}
\begin{tabular}{l c c c c}
\hline\hline
Architecture & Fermi-Hubbard & QPE-Abs & Stat-QPE & Stat-QPE-gap \\
\hline
\multicolumn{5}{l}{\textit{$p = 10^{-3}$}} \\
Baseline             & $4.85 \times 10^{13}$ & $4.20 \times 10^{13}$ & $4.91 \times 10^{10}$ & $1.69 \times 10^{10}$ \\
Compact              & $8.76 \times 10^{13}$ & $5.22 \times 10^{13}$ & $9.17 \times 10^{10}$ & $4.59 \times 10^{10}$ \\
AV                   & $4.60 \times 10^{12}$ & $1.20 \times 10^{12}$ & $9.51 \times 10^{9}$ & $3.27 \times 10^{9}$ \\
t-AV (atoms)         & $\mathbf{2.71 \times 10^{10}}$ & $\mathbf{4.58 \times 10^{10}}$ & $\mathbf{7.30 \times 10^{8}}$ & $\mathbf{2.21 \times 10^{8}}$ \\
t-AV (photonics)     & $1.94 \times 10^{11}$ & $2.63 \times 10^{11}$ & $2.03 \times 10^{9}$  & $7.62 \times 10^{8}$  \\
\hline
\multicolumn{5}{l}{\textit{$p = 10^{-4}$}} \\
Baseline             & $1.98 \times 10^{12}$ & $3.06 \times 10^{11}$ & $3.43 \times 10^{9}$  & $1.71 \times 10^{9}$  \\
Compact              & $6.27 \times 10^{12}$ & $1.04 \times 10^{12}$ & $1.36 \times 10^{10}$ & $6.81 \times 10^{9}$  \\
AV                   & $6.16 \times 10^{11}$ & $5.92 \times 10^{10}$ & $6.64 \times 10^{8}$ & $3.32 \times 10^{8}$ \\
t-AV (atoms)         & $\mathbf{1.10 \times 10^{10}}$ & $\mathbf{1.85 \times 10^{10}}$ & $\mathbf{2.25 \times 10^{8}}$ & $\mathbf{1.13 \times 10^{8}}$ \\
t-AV (photonics)     & $1.81 \times 10^{10}$ & $3.07 \times 10^{10}$ & $4.42 \times 10^{8}$  & $2.21 \times 10^{8}$  \\
\hline\hline
\end{tabular}
\end{table*}

\section{Conclusion}
\label{sec:conclusion}
In this work, we introduced a platform-aware compilation framework that maps a fault-tolerant algorithm, expressed as a logical Clifford+T quantum circuit, onto an instruction set matched to the connectivity classes of the target hardware platform. We considered three hardware connectivity classes: nearest neighbor (local), limited non-local, and effectively all-to-all, focusing particularly on superconducting, photonics, and neutral atoms. For each connectivity class, we outlined corresponding compilation architectures and described in detail how an algorithm can be efficiently compiled under the different architecture-hardware configurations. This allowed us to produce end-to-end resource estimates within a single pipeline, covering important quantities such as physical qubit count (including both data qubits and qubits for magic state factories), bridge-qubit demand for parallelization, as well as reaction depth across different hardware-architecture configurations. 

Building on the active-volume (AV) architecture, we introduced the transversal active volume (t-AV) architecture for platforms that support transversal logical Clifford operations. The t-AV architecture is designed for efficient execution of fault-tolerant operations on hardware with long-range connectivity, and we describe how it can be realized on both neutral atoms and photonic hardware platforms. We combined the t-AV architecture with four magic state factories: transversal distillation, distillation with a parity ancilla, lattice-surgery-based distillation, and fold-transversal cultivation. Furthermore, we described how these protocols can be combined with zero-level distillation to reduce space-time cost further.

We applied our framework to two benchmarks: Trotterized phase estimation of the low-lying eigenenergies of a small molecule (TMM), and a single fourth-order Trotter step of a $10\times10$ square lattice Fermi--Hubbard model (200-qubit system). 
Our TMM benchmark computation can be performed with $\sim 10^4$ physical qubits with the t-AV architecture, placing it within reach of early fault-tolerant demonstration of a quantum chemistry calculation, and with runtimes between $10^2$ ms (photonics, superconducting) and $ 10^5$ ms (neutral atoms). 
While the Fermi-Hubbard computation can be performed with $\sim 10^6$ physical qubits with a similar runtime to the TMM benchmark. 

We found that no single architecture is simultaneously optimal in space and time; each instead occupies a distinct region of the runtime-qubit resources plane. 
Our analysis also highlights trade-offs between runtime and the additional physical qubits needed to parallelize operations through increasing the number of magic state factories and bridge qubits. 
Neutral-atom t-AV achieves the smallest physical-qubit footprint and the lowest spacetime volume across every benchmark and error rate (Table~\ref{tab:stv-summary}), but pays the cost of a slow code cycle. 
Photonic t-AV achieves the shortest runtimes while reducing the physical-qubit count by roughly an order of magnitude relative to photonic AV, and the lattice-surgery baselines remain the most expensive.
The advantage of t-AV over AV is rooted in more efficient distillation: transversal Cliffords complete in $\mathcal{O}(1/d)$ logical cycles, meaning t-AV distills magic states relatively faster and with a space-time volume that scales as $d^2$.  
An analysis of reaction time and depth showed the runtime overhead from feedforward-limited measurement is negligible for t-AV on neutral atoms and modest for photonics.

These results show that the most efficient compilation paradigm is hardware-dependent, and that exposing this dependence, rather than committing to a single architectural model in advance, is essential for accurate cross-platform resource and performance comparisons. 
The framework is flexible and can be extended to include additional QEC codes, other distillation and cultivation protocols, different scheduling strategies, and other target algorithms. As fault-tolerant hardware matures, we expect platform-aware compilation to become essential for assessing the feasibility of running fault-tolerant quantum algorithms.

\section*{Code availability}
\label{code}
All code used to generate the results presented in this work, and the code to compile the respective circuits for each architecture considered in the work, is publicly available in the GitHub repository associated with the manuscript~\url{https://github.com/NQCP/NQCP-AA-git-zx-compiler}

\section*{AI use statement}
The scientific ideas reported in the manuscript originated from and were developed by the authors. AI tools (Claude) were used to assist in the code development for numerical simulations and plotting.  All hand-drawn figures were hand-drawn by a human.

\begin{acknowledgments}
This work is supported by the Novo Nordisk Foundation, Grant number NNF22SA0081175, NNF Quantum Computing Programme. S.X.C. acknowledges support from UK EPSRC (EP/SO23607/1). We thank Nick S. Blunt for valuable discussions on statistical phase estimation.
We also thank valuable discussions with Marcel David Fabian on making the compiler scalable. 
We thank Aliki Capatos, Ben Graham, Eoghan Ryan, Andreas B. Michelsen, Martin Hayhurst Appel, and Svend Krøjer Møller for useful discussions on different hardware platforms as well as active volume architecture. 

\end{acknowledgments}

\bibliography{main}

\appendix
\section*{Appendix}

\section{Pauli based Computation}
In this section, we briefly describe Pauli-based computation, which relies on Pauli product measurements (PPMs) to perform non-Clifford Pauli product rotations (PPRs).
\label{app:pauli-based-computation}
\subsubsection{Performing $P_{\pi/8}$ PPRs}
\label{subsubapp:performing-ppr}
\begin{figure}[h]
\centering
\includegraphics[width=0.6\columnwidth]{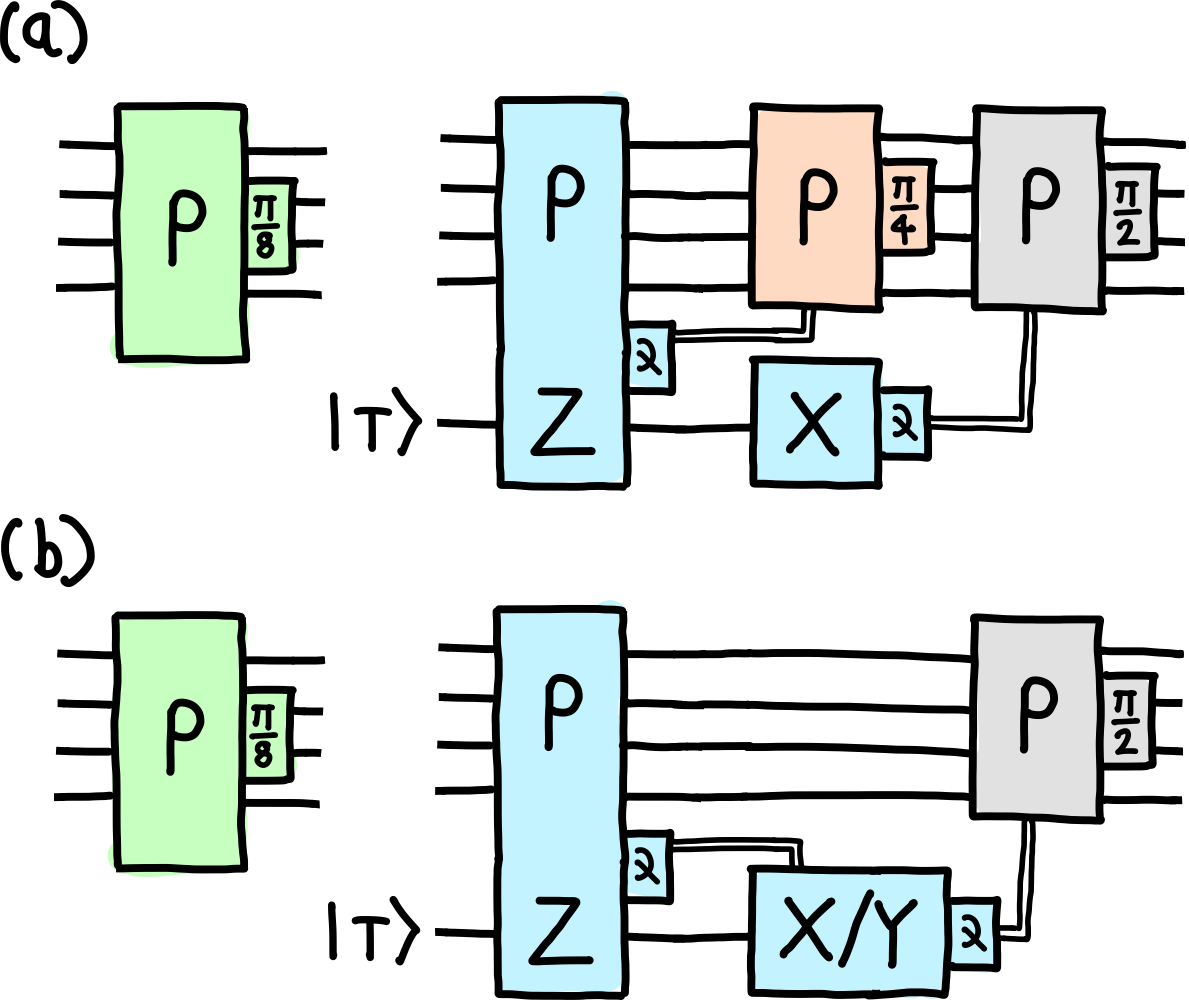}
\caption{(a) Performing $P_{\pi/8}$ PPR via joint $P \otimes Z$ measurement with a magic state. (b) auto-corrected PPRs from Ref. \cite{PBCGameSurface}}
\label{fig:ppr}
\end{figure}

A logical Clifford + T circuit can be recompiled in terms of $P_{\pi/8}$ PPRs ($P$ being a multi-qubit Pauli string) with Cliffords being absorbed into measurements at the end of the circuit\cite{PBCGameSurface}.
Each $P_{\pi/8}$ PPR can be performed by a joint Pauli $P \otimes Z$ measurement on $\text{data} \otimes \text{ancilla}$ register,  with ancilla being initialized in a high fidelity $\ket{T} = \ket{0} + i\pi/4\ket{1}$ state, followed by a $P_{\pi/4}$ Clifford correction on data qubits, see \figref{fig:ppr}{(a)}. 
The stale T state is then measured in the $X$ basis that generates a $P_{\pi/2}$ Pauli correction on data qubits that can be tracked classically.
Clifford corrections are too complex to track classically, so they must be applied directly, which makes in-line corrections a bottleneck in PBC, idling many qubits.
One way to overcome this bottleneck is to shift the corrections from PPR qubits to magic state qubits ~\cite{PBCGameSurface}.
The Clifford correction choice is deferred by encoding it into a later measurement basis choice on a \textit{stale} $T$ state.
As shown in \figref{fig:ppr}{(b)}, the PPR can be performed via such an auto-corrected implementation where corrections are moved away from data qubits to magic state qubits, and the measurement basis is adaptively chosen depending on the outcome of the $P \otimes Z$ measurement.

For example, a three-qubit PPR $(\mathrm{X}_1 \otimes \mathrm{Z}_2 \otimes \mathrm{X}_3)_{\pi/8}$ PPR is implemented by initializing a surface code patch in high fidelity $T$ state and performing a joint $(\mathrm{X}_1 \otimes \mathrm{Z}_2 \otimes \mathrm{X}_3 \otimes \mathrm{Z}_T)$ measurement between data patches and magic state patch.
Multi-Pauli measurements are performed via lattice surgery between data qubit patches and magic state patches ~\cite{SurfCodeLattice}.
To have a non-faulty PPR gadget, it requires the injection of high-quality magic states, as injecting noisy magic states will yield a faulty $P_{\pi/8}$ PPR.
There are several approaches for producing high-fidelity magic states, including distillation, state injection, and code switching ~\cite{MSDLow, TStateInjection, CodeSwitching}. 
Among these, magic state distillation has been studied most extensively, as explained in more detail in \ref{subapp:15-to-1-distillation}.
Since the measurement is non-destructive, the post-measurement T state needs to be removed via a destructive single-qubit measurement.
The basis of this measurement depends on its outcome ($P \otimes Z$ = +1, $X$ basis; $P \otimes Z$ = -1, $Y$ basis;). 
These measurements need to be performed within a code cycle. 
$Y$ measurements are not fast to perform within a code cycle on surface codes ~\cite{PBCGameSurface}. 
Thus, allowed measurements are single-qubit $X$ and $Z$ measurements and Bell-basis measurements $X \otimes X$ and $Z \otimes Z$.
Therefore, $Y$ measurements are performed by initializing a resource state $\ket{Y}$ followed by gate teleportation.

\section{Magic factories}
\label{app:magic-factories}
In this section, we describe the 15-to-1 magic state construction used to develop the various transversal variants in Sec. \ref{subsec:t-av-of-magic-factories-and-platform-specific-msd-layout-optimization}.
\subsection{15-to-1 distillation}
\label{subapp:15-to-1-distillation}

Magic state distillation refers to producing high-quality $T$ states by using many noisy copies of $T$ states.
The logical error rate of produced magic states typically scales polynomially with the physical error rate. 
A certain family of codes admitting a transversal $T$ gate ensures such error scaling up to some degree, the smallest instance being a $[[15, 1, 3]]$ quantum Reed-Muller code, where applying $T$ to all 15 physical qubits implements a logical $T$.
15-to-1 Magic state distillation protocol is based on this code, which takes 15 many noisy copies of $\ket{T}$ states as input to produce one high-fidelity magic state with an output error rate of the order of $\mathcal{O}(p^3)$.
Depending on the platform as well as code compatibility, $15\text{-to-}1$ admits different implementations. 
If the platform supports effectively all-to-all connectivity and transversal gates on the surface code, as shown in Fig.~\ref{fig:t-msd}, the 15-to-1 distillation protocol can be implemented as follows: 
We start by initializing the 15 logical qubits, each encoded in a surface code patch, into either $\ket{0}$ or $\ket{+}$. 
CNOT construction is the encoding circuit of the protocol, forming a Bell state of the 0-th qubit entangled with the Reed-Muller code qubits (qubits 1 to 15). 
All of the CNOTs are implemented transversely.  
After this, a logical T gate is performed by injecting 15 noisy T states via gate teleportation.

\begin{figure}[h]
\centering
\includegraphics[width=\columnwidth]{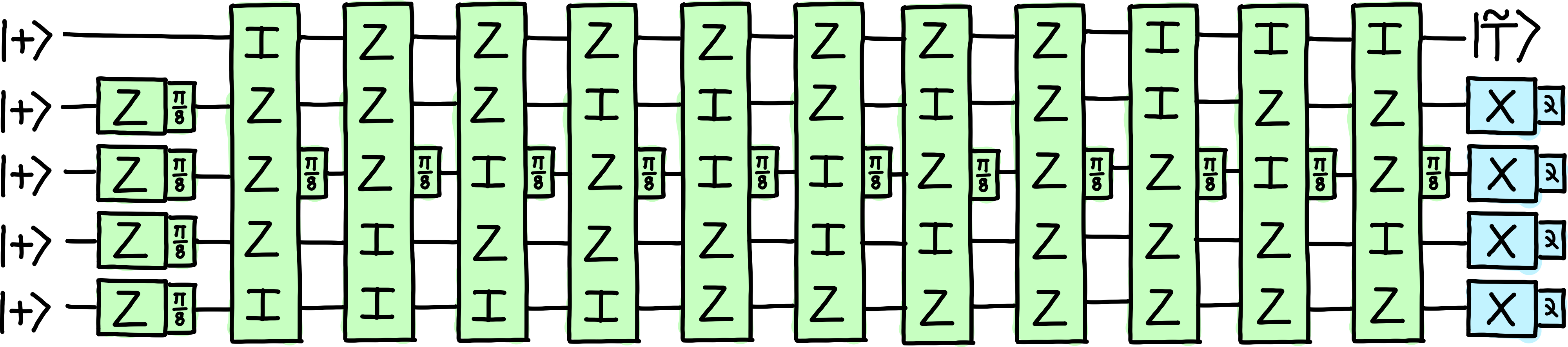}
\caption{Re-written 15-to-1 circuit in PBC}
\label{fig:15to1pbc}
\end{figure}

\begin{figure}
\centering
\includegraphics[width=1.0\columnwidth]{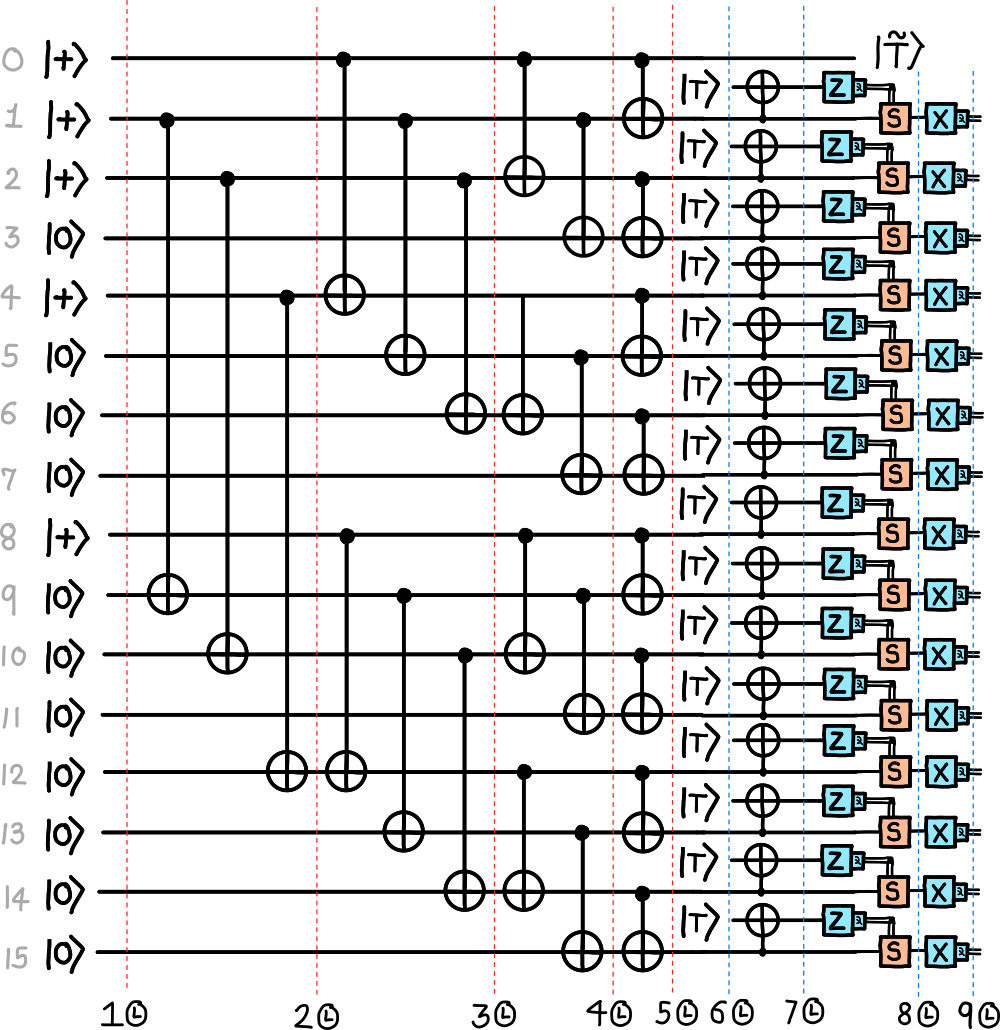}
\caption{CNOT construction of 15-to-1 magic state distillation protocol. Red and Blue lines indicate syndrome extraction cycles. Transversal implementation of the protocol requires $9 \clk_c$ in total.}
\label{fig:t-msd}
\end{figure}

On platforms with local and limited non-local connectivity, transversal implementation is not supported; hence, the protocol needs to be rewritten in terms of PPRs. 
As we saw in \ref{subsec:pauli-based-computing}, any logical circuit can be expressed in terms of $P_{\pi/8}$ rotations; certain sequences of $P_{\pi/8}$ PPRs are equivalent to the identity. 
The 15to1 PPR circuit builds on such sequences.
Fig.~\ref{fig:15to1pbc} shows a non-trivial example of 15 mutually commuting $P_{\pi/8}$ rotations that are equivalent to the identity. 
Such sequences are obtained using tri-orthogonal matrices and phase polynomials that always evaluate to zero ~\cite{PBCGameSurface}.
If such a circuit is initialized with five $\ket{+}$ states, it produces a magic state on the first qubit up to a Clifford correction, leaving the rest of the $\ket{+}$ states unchanged. 
To detect any errors happening on the remaining 4 qubits, they are measured in the $X$ basis; if any of the outcomes gives $-1$, the protocol is repeated.

\subsection{Fold transversal surface code cultivation}
\label{subapp:fold-transversal-surface-code-cultivation}
The magic-state cultivation~\cite{GidneyCultivation} protocol prepares high-quality magic states by post-selecting on the measurement of a transversal logical Clifford operator on an initial small-distance code and then rapidly growing to a larger-distance code.
Cultivation reaches comparable output fidelities at a significantly lower spacetime footprint than lattice-surgery-based magic state distillation~\cite{GidneyCultivation, Sahay2025FoldTransversal}. 
Fold-transversal cultivation on the surface code~\cite{Sahay2025FoldTransversal} attains the lowest known spacetime overhead to prepare a $\ket{T}$ state by measuring fold-transversal $H_{XY}(=(X+Y)/\sqrt{2})$ of the unrotated surface code, and is therefore well suited to platforms with reconfigurable connectivity such as atom arrays. 
Due to its attained lower error rates, this protocol has been used to perform end-to-end resource estimation of different early fault-tolerant algorithms on Neutral atoms~\cite{khan2026architectingearlyfaulttolerant}.
Note that the error rates achieved by the cultivation scheme are suitable for small-scale use cases, and one would need multiple layers (eg., distillation on top of the cultivation) to solve use cases demonstrating quantum advantage. 
Fold transversal cultivation with a fault distance of 5 under a uniform depolarizing noise model with a physical error rate of $p = 10^{-3}$ achieves a logical error rate of $10^{-8}$ with 5 expected attempts \cite{Sahay2025FoldTransversal}. 
Cultivation with fault distance 5 goes through the following stages: 
\begin{itemize}
    \item \textbf{Injection stage $(1 \clk_c)$}: is done via preparing a logical magic state $\ket{T}=\ket{0} + e^{i\pi/4}\ket{1}$ in a distance-3 rotated surface code using a unitary encoding circuit which takes 8 time steps, followed by a round of stabilizer measurements. 
    \item \textbf{Cultivation stage $(6 \clk_c)$}: The rotated surface code is transformed or morphed into a regular/unrotated surface code, which simply performs the first two time steps of the syndrome extraction circuit. 
    After this, the fold-transversal logical $H_{XY} = (X + Y)/\sqrt{2}$ operator is measured twice using a GHZ(3) ancilla state, which takes 2 code cycles. $(2 \clk_c)$
    After this, a unitary growth to the rotated surface code 5 is performed, followed by two rounds of stabilizer measurements. $(2 \clk_c)$.
    The code is again morphed into a regular surface code of distance 5, followed by $H_{XY}$ logical checks $(2 \clk_c)$.
    The code is then grown to a rotated surface code of distance 9.
    \item \textbf{Escape stage $(5 \clk_c)$:} In the escape stage, the code is grown to a larger distance, via stabilizer measurements, to a distance of 11. This takes 5 rounds of stabilizer measurements. $(5 \clk_c)$.
\end{itemize}
After this, a code has a magic state encoded with a logical error rate of $10^{-8}$ and is grown via unitary growth to match the distance of the data patch. 
Based on these numbers, we calculate the space footprint in terms of physical qubits of the cultivation protocol as $Q_{\mathrm{cult}}(d) = 562 + d_{\mathrm{eff}}^2$, with $d_{\mathrm{eff}} = \min(d, 11)$.
$562$ physical qubits is the fixed cultivation overhead and $d_{\mathrm{eff}}$ is the grown surface-code patch distance, capped at $11$ \cite{Sahay2025FoldTransversal}.
The factory operates on a per-attempt schedule of duration $r_{\mathrm{att}} = 12 \clk_c$, assuming neutral atoms code cycle as $10\,\mathrm{ms}$, this corresponds to  $t_{\mathrm{att}} = 120\,\mathrm{ms}$.
Due to post-selection, each cultivation attempt succeeds only with probability $1 - \delta_r$, where $\delta_r$ is the protocol's \emph{discard rate}. 
The expected number of attempts per produced $\ket{T}$ state is therefore $ n_{\mathrm{att}} = 1/(1 - \delta_r)$, and the effective cost in code cycles per state ($n_{\text{cyc}}$) is
\begin{equation}
    n_{\mathrm{cyc}}(p_{\mathrm{phys}}) \;=\; r_{\mathrm{att}} \cdot  n_{\mathrm{att}}(p_{\mathrm{phys}}) \;=\; \frac{r_{\mathrm{att}}}{1 - \delta_r(p_{\mathrm{phys}})}.
    \label{eq:cult-cycles-model}
\end{equation}

The discard rate is dominated by single-fault rejections during the post-selected stages, so it is well described by a Poisson model:
\begin{equation}
    1 - \delta_r(p_{\mathrm{phys}}) \;=\; \exp\!\bigl(-N\,p_{\mathrm{phys}}\bigr),
    \label{eq:cult-poisson}
\end{equation}
where $N$ is the effective number of error-sensitive locations per attempt. Both~\cite{Sahay2025FoldTransversal} and~\cite{khan2026architectingearlyfaulttolerant} report all their cultivation numbers at $p_{\mathrm{phys}} = 10^{-3}$; we calibrate $N$ to their reported numbers and extrapolate to $p_{\mathrm{phys}} = 10^{-4}$ using the same model.
With $\delta_r = 0.80$ at $p_{\mathrm{phys}} = 10^{-3}$, we get, $n_{\mathrm{att}} = 5$, and $n_{\mathrm{cyc}}(10^{-3}) = 60 \clk_c$ per cultivated state.
This sets the Poisson normalization in Eq.~\eqref{eq:cult-poisson} to $N \approx -\ln(0.2)/10^{-3} \approx 1609$.
Substituting the calibrated $N$ into Eq.~\eqref{eq:cult-poisson} at $p_{\mathrm{phys}} = 10^{-4}$ gives $\delta_r \approx 1 - \exp(-0.161) \approx 0.149$, so $n_{\mathrm{att}} \approx 1.18$. Rounded conservatively upward to $ n_{\mathrm{att}} = 1.5$, giving:
\begin{equation}
    n_{\mathrm{cyc}}\bigl(10^{-4}\bigr) \;=\; 12 \times 1.5 \;=\; 18.0 \; \clk_c.
    \label{eq:cult-1e-4}
\end{equation}
Eq.~\eqref{eq:cult-1e-4} is an analytical extrapolation; it is conservative relative to the unrounded Poisson estimate ($\sim 14$ cycles/state) and our cult-t-AV curves at $p_{\mathrm{phys}} = 10^{-4}$ should be read as an upper bound on cultivation cost.

\begin{table*}[htbp]
\centering
\caption{Output logical error probabilities $p_{out}$ of the magic-state protocols considered in this work, at physical error rates $p = 10^{-3}$ and $10^{-4}$, with their corresponding space-time . Concat LS-dist numbers are taken from code provided in Ref. \cite{MSDNotCostly}.}
\label{tab:15-to-1-comparison}
\begin{tabular}{l c c c c}
\hline\hline
Magic factory variant & Protocol  & $p$  & $p_{out}$ & Physical STV \\
\hline
trans-dist & $(15\text{-to-}1)_{d =19/17/15/13}$   & $10^{-3}$  & $3.5 \times 10^{-8}$ & $267d^2$\\
& $(15\text{-to-}1)_{d = 9/7}$   &  $10^{-4}$ & $3.5 \times 10^{-11}$ & \\
\hline
trans-dist (erasure) & $(15\text{-to-}1)_{d = 11/9}$   &  $10^{-3}$ & $3.5 \times 10^{-11}$ & $267d^2$\\
& $(15\text{-to-}1)_{d = 7/5}$   &  $10^{-4}$ & $3.5 \times 10^{-14}$ & \\
\hline
parity-dist & $(15\text{-to-}1)_{d =19/17/15/13}$   & $10^{-3}$  & $3.5 \times 10^{-8}$ & $217d^2$\\
& $(15\text{-to-}1)_{d = 9/7}$   &  $10^{-4}$ & $3.5 \times 10^{-11}$ & \\
\hline
parity-dist (erasure) & $(15\text{-to-}1)_{d = 11/9}$   &  $10^{-3}$ & $3.5 \times 10^{-11}$ & $217d^2$\\
& $(15\text{-to-}1)_{d = 7/5}$   &  $10^{-4}$ & $3.5 \times 10^{-14}$ & \\
\hline
LS-dist (NN) & $(15\text{-to-}1)_{d =23/21/19/17/15}$~\cite{PBCGameSurface}   & $10^{-3}$  & $3.5 \times 10^{-8}$ & $242d^3$\\
& $(15\text{-to-}1)_{d = 11/9/7}$   &  $10^{-4}$ & $3.5 \times 10^{-11}$ & \\
LS-dist (logN) & $(15\text{-to-}1)_{d_X =21/19/17/15, d_Z = d_m =9/7}$~\cite{AVLitinski}   & $10^{-3}$  & $1 \times 10^{-8}$ & $35d^3$\\
& $(15\text{-to-}1)_{d = 11/9/7, d_X = d_m = 5/3}$   &  $10^{-4}$ & $1.2 \times 10^{-11}$ & \\
\hline
0-dist + trans-dist & $(15\text{-to-}1)_{d =19/17}$   & $10^{-3}$  & $3.5 \times 10^{-11}$ & $657d^2$ \\
0-dist + parity-dist & $(15\text{-to-}1)_{d =19/17}$   & $10^{-3}$  & $3.5 \times 10^{-11}$ & $607d^2$\\
0-dist + LS-dist & $(15\text{-to-}1)_{d =19/17}$ & $10^{-3}$  & $3.5 \times 10^{-11}$ & $390d^2 + 43d^3$\\
\hline
concat LS-dist (logN) & $(15\text{-to-}1)_{d/2, d/4, d/4}\times(8\text{-to-CCZ})_{d, d, d/2}$~\cite{AVLitinski}  & $10^{-3}$  & $3.5 \times 10^{-11}$ & $51.5d^3$\\
\hline
concat LS-dist (NN) & $(15\text{-to-}1)^6_{d_X = 11, d_Z = 5, d_m = 5}\times 8\text{-to-CCZ})_{d_X = 21, d_Z = 13, d_m = 13}$~\cite{MSDNotCostly}  & $10^{-3}$  & $5.9 \times 10^{-10}$ & $32242 \times 52$\\
concat LS-dist (NN) & $(15\text{-to-}1)^6_{d_X = 11, d_Z = 5, d_m = 5}\times 8\text{-to-CCZ})_{d_X = 19, d_Z = 13, d_m = 13}$~\cite{MSDNotCostly}  & $10^{-3}$  & $5.2 \times 10^{-9}$ & $30542 \times 52$\\
concat LS-dist (NN) & $(15\text{-to-}1)^6_{d_X = 11, d_Z = 5, d_m = 5}\times 8\text{-to-CCZ})_{d_X = 23, d_Z = 13, d_m = 13}$~\cite{MSDNotCostly}  & $10^{-3}$  & $1.4 \times 10^{-10}$ & $34086 \times 52$\\
\hline
cultivation & fold-transversal~\cite{Sahay2025FoldTransversal} & $10^{-3}$ & $10^{-8}$ & $(562 + d_{\text{eff}}^2)\cdot60$\\
& fold-transversal & $10^{-4}$ & $ < 10^{-8}$ & $(562 + d_{\text{eff}}^2)\cdot18$\\
cultivation (erasure) & fold-transversal & $10^{-3}$ & $10^{-11}$ & $(562 + d_{\text{eff}}^2)\cdot60$\\
& fold-transversal & $10^{-4}$ & $< 10^{-11}$ & $(562 + d_{\text{eff}}^2)\cdot18$\\

\hline\hline
\end{tabular}
\end{table*}

\section{Fault-tolerant Architectures}
\label{app:architectures}
\subsection{Baseline architecture}
\label{subapp:baseline-architecture}
In baseline architectures, a logical data qubit is encoded in a rectangular surface code patch of size $2d \times d$, requiring $4d^2$ physical qubits. 
Logical qubits are arranged in a 2D static layout with nearest-neighbor connectivity, with data qubits lying side by side in a row.  
Such a layout gives direct access to $X$, $Y$, and $Z$ boundaries of a surface code in any given logical cycle (See Fig. 1 (c) of Ref. \cite{PBCGameSurface}).
If one has direct access to all the Pauli boundaries, PPRs can be performed in $1 \, \clk$ via lattice surgery, given that a magic state factory produces $T$ states to match this consumption rate.
The execution of such a PPR is mediated by initializing an ancillary region spanning the logical data qubits row, wide enough to perform any multi-qubit PPR between arbitrary logical data qubits in a single code cycle. 
In such a setup, PPRs are executed strictly sequentially at a rate of one PPR per logical cycle, and the resulting space-time volume $V_{\text{base}}$ of the computation is given by:
\begin{equation}
    V_{\mathrm{base}} \;=\; 4 \cdot n_Q \cdot n_T,
\end{equation}
where $n_T$ is the number of non-Clifford gates ($T$ gates) in the logical circuit.
Such construction incurs a substantial amount of idle volume due to the ancillary region used for lattice surgery. 
For instance, if one of the PPRs contains PPM of the form $I_1 \otimes Z_2 \otimes I_3 \otimes ... \otimes I_{n_Q}$, during computation, only one data patch takes part in the active computation; the remaining $n_Q - 1$ patches remain idle but incur the same error correction overhead as the active data patch.  
This large idle volume motivates the compact architecture design explained below and the active-volume architecture described in subsequent sections.

\subsection{Compact architecture}
\label{subapp:compact-architecture}
The compact architecture reduces the data-block footprint by encoding the logical qubit in a single surface code patch, which subsequently reduces the footprint of the workspace ancilla ~\cite{PBCGameSurface}.
In the compact setup, not all boundaries are accessible at a given logical cycle. 
If the PPR contains $X$ measurement, the patch needs to be rotated by $\pi/2$, which takes $3 \, \clk$ ~\cite{PBCGameSurface}. 
Since $Y$ boundaries are not accessible even with rotation, if PPR contains an odd number of $Y$ operators, it can be decomposed into PPRs with only $X$ and $Z$ operators by introducing $P_{\pi/4}$ rotations as corrections. 
For example, a PPR of the form $(X_1 \otimes Y_2 \otimes Z_3 \otimes Z_4)_{\pi/8}$ can be decomposed as $(I_1 \otimes Z_2 \otimes I_3 \otimes I_4)_{\pi/4}(X_1 \otimes X_2 \otimes Z_3 \otimes Z_4)_{\pi/8}(I_1 \otimes Z_2 \otimes I_3 \otimes I_4)_{-\pi/4}$.
The last $(I_1 \otimes Z_2 \otimes I_3 \otimes I_4)_{-\pi/4}$ rotation can be commuted past and absorbed into measurements in the end; however, we need to perform the first $(I_1 \otimes Z_2 \otimes I_3 \otimes I_4)_{\pi/4}$ rotation explicitly by doing a joint $I_1 \otimes Z_2 \otimes I_3 \otimes I_4 \otimes Y$ measurement using an extra ancilla that has accessible $Y$ boundaries. 
Such $P_{\pi/4}$ rotation involves one PPM on $Z$ boundaries and hence takes $1\,\clk$, therefore, in total, PPR with odd number of $Y$ operators can take upto $4\,\clk$.
The most time-consuming PPR is one containing an even number of $Y$ operators; it needs to be decomposed into two explicit $P_{\pi/4}$ rotations to avoid negative global phase on Pauli operators that constitute the corresponding PPR.
For example, PPR of type $(X_1 \otimes Y_2 \otimes Z_3 \otimes Y_4)_{\pi/8}$ needs to be decomposed into $(I_1 \otimes Z_2)_{\pi/4}(I_3 \otimes Z_4)_{\pi/4}(X_1 \otimes X_2 \otimes Z_3 \otimes X_4)_{\pi/8}(I_1 \otimes Z_2)_{-\pi/4}(I_3 \otimes Z_4)_{-\pi/4}$. 
Here, we need to perform the first two explicitly, which takes $2\,\clk$, and hence the PPR can take up to $9 \,\clk$.
Therefore, the consumption of magic state on a compact block can take up to $9 \, \clk$, and the total qubit footprint for $n_Q$ logical qubits is $1.5\, n_Q + 3$ patches.

The $15\text{-to-}1$ distillation circuit of Fig.~\ref{fig:15to1pbc} can be laid out using $11$ patches, and as it contains $11$ PPRs containing only $Z$ operators, the whole distillation protocol is completed in $11\,\clk$, so the bottleneck is now magic-state production rather than PPR consumption.
This yields a longer runtime than the baseline (by a factor of $\sim 11$) in exchange for a reduced spatial footprint and idle volume, making the compact model preferable at fixed qubit budgets.
The total space-time volume becomes:
\begin{equation}
    V_{comp} = (1.5\, n_Q + 14)\cdot 11\cdot n_T 
\end{equation}
\textbf{Space-time cost:} Such architectures have a memory space cost of $1.5\, n_Q + 3$ logical patches and distillation space cost of $11$ logical patches and time cost of $11 \cdot n_T \, \clk$.

\subsection{Active volume architecture}
\label{subapp:active-volume-architecture}
In this section, we give a detailed overview of the active volume (AV) architecture considered in this work.
AV architecture is a lattice-surgery-based architecture suitable for platforms supporting limited non-local connectivity between physical qubits, such as photonics\cite{AVLitinski}. 
Such an architecture partitions the quantum computer into two distinct zones: a memory zone and a workspace zone. 
Memory qubits are stored in a memory zone, rearranged beforehand for the subsequent logical operations. 
Qubits that are required to be measured after the computation are also stored here. 
A workspace zone is dedicated to carrying out logical operations.
Whenever a scheduled operation contains some of the logical qubits stored in a memory, they are sent to the workspace zone via transversal SWAPs ($0 \clk$). 
Distillation of magic states also happens in the workspace. 
The architecture supports PPR/PPM-type compilation via lattice surgery, but PPMs do not map naturally as directly implementable objects on the platform.
It is useful to re-express such PPMs in terms of their equivalent ZX diagrams; further imposing a set of rules on these ZX diagrams leads us to oriented ZX diagrams, which are famously called logical blocks in the AV architecture. 
Logical blocks follow a certain set of rules, which can be summarized as follows:
\begin{itemize}
    \item \textbf{Type} a logical block is a ZX spider with at most four legs, carrying a type (Z: Orange or X: Blue) 
    \item \textbf{Orientation} The orientation (E, N, or U) of the block is the axis along which the block has no ports. 
    A U-oriented block is "temporally flat" (no U/D ports), which is a spatial-only operation, e.g., ancillas mediating PPMs.
    \item \textbf{Hadamarded ports} A block has at most one input leg, always in the D direction. A port can be Hadamarded, which applies a Hadamard to the qubit on that port.
    \item \textbf{Commensurability}: Two blocks are commensurate if they share both type and orientation, or differ in both; otherwise, they are incommensurate.
    \item \textbf{Connection rule} A port connection implements a Bell-state projection $(|00\rangle + |11\rangle)/\sqrt{2}$ between the two qubits involved. 
    Connections are only allowed between ports pointing in the same direction. 
    Two connected blocks must be commensurate, unless exactly one of the two ports is Hadamarded, in which case they must be incommensurate. 
    In a completed logical operation, all W/S/E/N ports are connected; unconnected D-ports are inputs, and unconnected U-ports are outputs.
\end{itemize}

Using the above rules, one can construct a network of logical blocks that represent PPMs/PPRs. 
As an example, we saw in Fig.~\ref{fig:monsterfigure}(b) that a PPM can be expressed in terms of a network of two logical blocks.
Below, we describe how to construct such a logical block network for any PPR/PPM.

\subsubsection{PPR compilation into logical network diagrams}
\begin{figure*}
\centering
\includegraphics[width=0.7\linewidth]{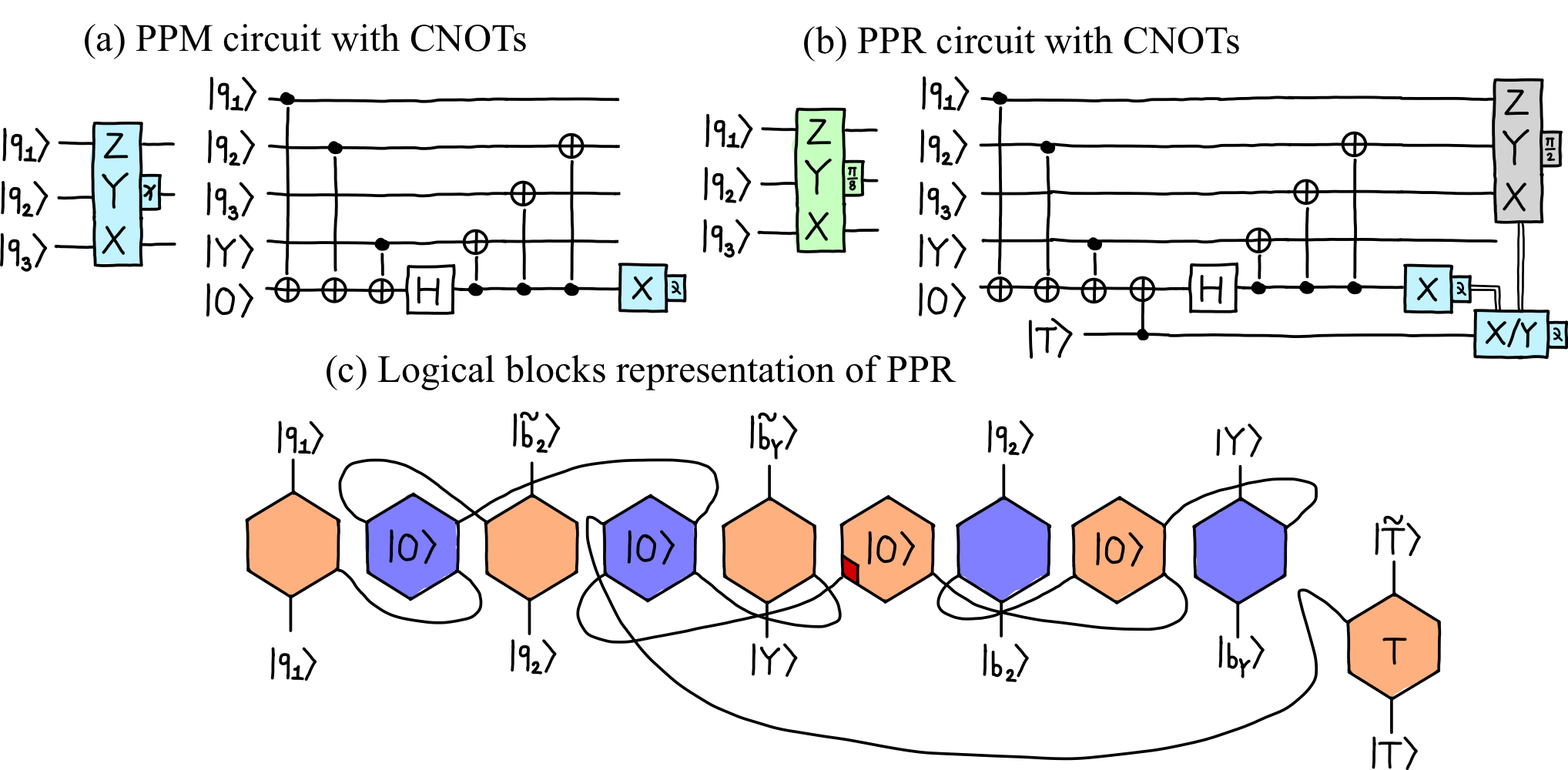}
\caption{(a) a CNOT circuit representation of $Z \otimes Y \otimes X$ PPM. (b) A CNOT circuit representation of PPR with $T$ state injection. (c) Logical block representation of the considered PPR}
\label{fig:logical-blocks}
\end{figure*}

As described in section \ref{app:pauli-based-computation}, PPRs of the form $P_{\frac{\pi}{8}} = e^{iP\frac{\pi}{8}}$ can be performed by consuming a magic state via a joint PPM between PPR qubits and a $\ket{T}$ state qubit.
Logical blocks construction of PPRs can be understood through their logical circuit. 
An arbitrary PPM containing $\mathrm{X},\mathrm{Y}$, and $\mathrm{Z}$ operators can be decomposed into a series of CNOTs and a parity ancilla qubit initialized in state $\ket{0}$.
The computation can be seen as two parts separated by a Hadamard: the first half of the computation contains a series of CNOTs with control qubits being the ones contributing to $\mathrm{Z}$ and $\mathrm{Y}$ operators in PPM and the ancilla as a target, followed by a basis change of the parity ancilla with $\mathrm{H}$.
The second half contains a series of CNOTs with the control being the ancilla and the target on qubits contributing to $\mathrm{X}$ and $\mathrm{Y}$ operators in PPM, followed by an $\mathrm{X}$ measurement on the ancilla. 
This circuit works when the PPM has an even number of $\mathrm{Y}$ operators; for PPMs with an odd number of $\mathrm{Y}$ operators, a resource state $\ket{\mathrm{Y}} = (\ket{0} + i\ket{1})/\sqrt{2}$ needs to be initialized on a fresh ancilla qubit, which acts as a control.
A corresponding CNOT circuit for $(Z \otimes Y \otimes X)_{\pi/8}$ PPM is shown in Fig.~\ref{fig:logical-blocks}\textcolor{blue}{(a)}.
Qubits carrying $\mathrm{Y}$ operators (and the $\ket{\mathrm{Y}}$ resource state, when present) participate in both the $Z$-half and the $X$-half of the PPM decomposition; therefore, in the logical-block network, they appear in both halves, which means they need to be used twice within a single cycle.
In such cases, bridge qubits (as described in Sec. \ref{subsec:parallelization-and-bridge-qubit-demand}) are initialized in the memory zone to teleport the corresponding qubit in the other half of the computation. 

For example, as shown in Fig.~\ref{fig:logical-blocks}\textcolor{blue}{(b)}, a $(Z \otimes Y \otimes X)_{\pi/8}$ PPR is first decomposed into a corresponding CNOT circuit and subsequently into a logical block network (Fig.~\ref{fig:logical-blocks}\textcolor{blue}{(c)}) as follows: to implement a PPM, an extra parity ancilla initialized in a $\ket{0}$ state is needed, in a logical network, the ancilla can have at most 4 port connections (two in N-S direction and two in E-W direction).
Out of these, only N-S can be used to connect the PPM qubits due to commensurability. 
The E-W direction is reserved as an extended connection of the ancilla $\ket{0}$. 
Therefore, if the ancilla $\ket{0}$ is connected to $w_{Z/Y}$ number of PPR qubits that contribute to Pauli $Z$ and $Y$ operations in the first half, and $w_{X/Y}$ number of PPR qubits that contribute to Pauli $X$ and $Y$ operations in the second half, one needs $\lfloor({w_{ZY} + 1})/2\rfloor$ logical blocks for the ancilla in the first half and $\lfloor({w_{XY} + 1})/2\rfloor$ in the second half.
Summing the two halves with the connector block that joins them yields a closed-form active volume for a weight-$(w_X, w_Z)$ PPM,
\begin{equation}
    AV_{\mathrm{PPM}(w_X, w_Z)} \;=\; \lceil 3 w_X / 2 \rceil + \lceil 3 w_Z / 2 \rceil + 1,
    \label{eq:av-ppm}
\end{equation}
where $w_X$ counts qubits carrying $X$ or $Y$, $w_Z$ counts qubits carrying $Z$ or $Y$, and both are incremented by one when the total $Y$ count is odd to account for the $\ket{\mathrm{Y}}$ resource state. 
Given a logical circuit, our hardware-informed circuit compiler produces these logical networks per PPR as hexagon objects with corresponding port connections, and code is openly available at \url{https://github.com/NQCP/NQCP-AA-git-zx-compiler}.
Logical blocks are executed in the workspace, while the memory module stores data qubits, resource states such as $\ket{Y}$, stale or surplus $\ket{T}$ states supplied by the distillation block, and bridge qubits.

\subsubsection{Magic state distillation in AV architecture}
The spatial overhead of $15$-to-$1$ distillation protocol of Fig.~\ref{fig:15to1pbc} can be further reduced by halving the measurement distance $d_m$  and Z boundary distance $d_Z$, with resulting $(15\text{-to-}1)_{d, d/2, d/2}$ protocol ~\cite{MSDNotCostly}.
This protocol can be laid out in the AV architecture as a network of $35$ half logical blocks~\cite{AVLitinski}.
With $p_{in}$ being the state-injection error rate and $p(d) = 10^{-d/2}$ the per-spacetime-block surface-code error rate at $10\%$ of the threshold, the rough analytical estimate of the output error rate of a variant $(15\text{-to-}1)_{d_X, d_Z, d_m}$ is
\begin{equation}
    p_{out} \;\approx\; 35 \cdot \bigl[4\,p(d/4) + p_{in}\bigr]^3 \;+\; 2\,p(d_X),
    \label{eq:15to1-error-av}
\end{equation}
The standard variant $(15\text{-to-}1)_{d, d/2, d/2}$ uses $35$ half-distance blocks with an active volume of $17.5$ blocks per $T$ state, producing $2$ $T$ states per logical cycle. 
Halving all distances to $(15\text{-to-}1)_{d/2, d/4, d/4}$ allows four instances to run simultaneously, one per quadrant of the workspace, giving us $16$ $T$ states per logical cycle from the same $35$ workspace qubits. 
$15\text{-to-}1$ alone cannot achieve the desired error rate; therefore, one needs to use concatenated protocols to produce magic states with low fidelity. 
Therefore, the generated magic states from $(15\text{-to-}1)_{d/2, d/4, d/4}$ are further fed to $(8\text{-to-CCZ})_{d, d, d/2}$ distillation protocol ~\cite{AVLitinski}.
The output error rate of $(8\text{-to-CCZ})_{d, d, d/2}$ is 
\begin{equation}
    p_{out} \;\approx\; 28 \cdot \bigl[4\,p(d/2) + p_{in}\bigr]^2 \;+\; 2\,p(d).
    \label{eq:8toCCZ-error}
\end{equation}
This second stage of distillation produces CCZ states and uses an additional 25 logical blocks.
Therefore, the active volume of the concatenated protocol becomes 30.
Some extra volume is required for state injection, so on average, the cost of a CCZ state is 35.
A distilled CCZ can be converted into two T states with an extra cost of 16.5 logical blocks\cite{AVLitinski,Gidney2019efficientmagicstate}. 
Therefore cost per $T$ state is $(16.5 + 35)/2 = 51.5/2 = 25.75$ logical blocks. 
The effective space-time volume of the concatenated protocol is therefore $25.75 \times 2d^3 = 51.5\,d^3$ physical qubit$\cdot$code-cycles per $T$ state.

\subsubsection{Reaction depth in AV architecture}
Similar to t-AV, computation in AV architecture can become reaction-limited. 
A new reaction depth layer is added whenever the pending Pauli correction from a previous reactive measurement anticommutes with the basis of the next reactive measurement on the same qubit; commuting corrections can be absorbed into the Pauli frame and do not require a new decoder round.
A logical cycle of reaction depth $k_{max}$ therefore requires $k_{max}\,\tau_r$ time of classical processing, which must fit within the physical duration $d\,\tau_c$ of one logical cycle.
When $k_{max}\,\tau_r > d\,\tau_c$, the processor must insert idle code cycles between reactive measurements; we refer to this excess as \textit{stalling}.
In the AV architecture, two anticommuting PPRs on overlapping qubits can be parallelized within a single logical cycle by introducing a bridge qubit, which teleports the data qubit between the two PPMs via a Bell measurement.
The Bell measurement basis is fixed ($XX$, $ZZ$), so it does not by itself add a reaction layer.
However, the Pauli correction produced by the first PPR propagates onto the teleported qubit and, by construction, anticommutes with the basis of the second PPR; the second PPM is therefore reactive on the first, contributing one additional layer.
A chain of $n$ pairwise-anticommuting PPRs within a logical cycle accordingly has reaction depth $n$, matching the value for the Solovay--Kitaev decomposition of arbitrary-angle PPRs~\cite{AVLitinski}.

\section{Quantum simulation problems \label{app:quantum-simulation-problems}}
We apply our compilation framework to two quantum simulation problems. First, we consider the calculation of eigenenergies of a small chemical system using Trotterized quantum phase estimation. Second, we compile a single 4th-order Trotter step of a $10\times 10$ square lattice Hubbard model. The Clifford+T circuits (the input circuits to our compilation code) for these problems are available on the GitHub repository. In Table~\ref{tab:msd-requirements}, we summarize the number of required Trotter steps ($r$), total T gate cost ($n_T$), and the magic-state error-rate requirement ($p_{\mathrm{out}}$) of each application considered in this paper. We describe these problems and the construction of their respective logical Clifford+T circuits below.

\begin{table}[tbp]
\centering
\begin{tabular}{l c c c}
\hline\hline
Benchmark & $r$ & $n_T$ & $p_{out} \le \epsilon/n_T$ \\
\hline
QPE-Abs       & $1680$ & $1{,}370{,}880$ & $7.3 \times 10^{-9}$ \\
Stat-QPE      & $40$   & $32{,}480$      & $3.1 \times 10^{-7}$ \\
Stat-QPE-gap  & $20$   & $16{,}240$      & $6.2 \times 10^{-7}$ \\
Fermi-Hubbard & $1$    & $805{,}804$     & $1.2 \times 10^{-8}$ \\
\hline\hline
\end{tabular}
\caption{Total $T$ count $n_T$ of each benchmark and the resulting magic-state
error-rate requirement $p_{out} \le \epsilon/n_T$ at the target failure budget
$\epsilon = 0.01$.}
\label{tab:msd-requirements}
\end{table}

\subsection{Trotterized quantum phase estimation \label{subapp:trotterized-quantum-phase-estimation}}

\begin{figure}[t]
\includegraphics[width=\columnwidth]{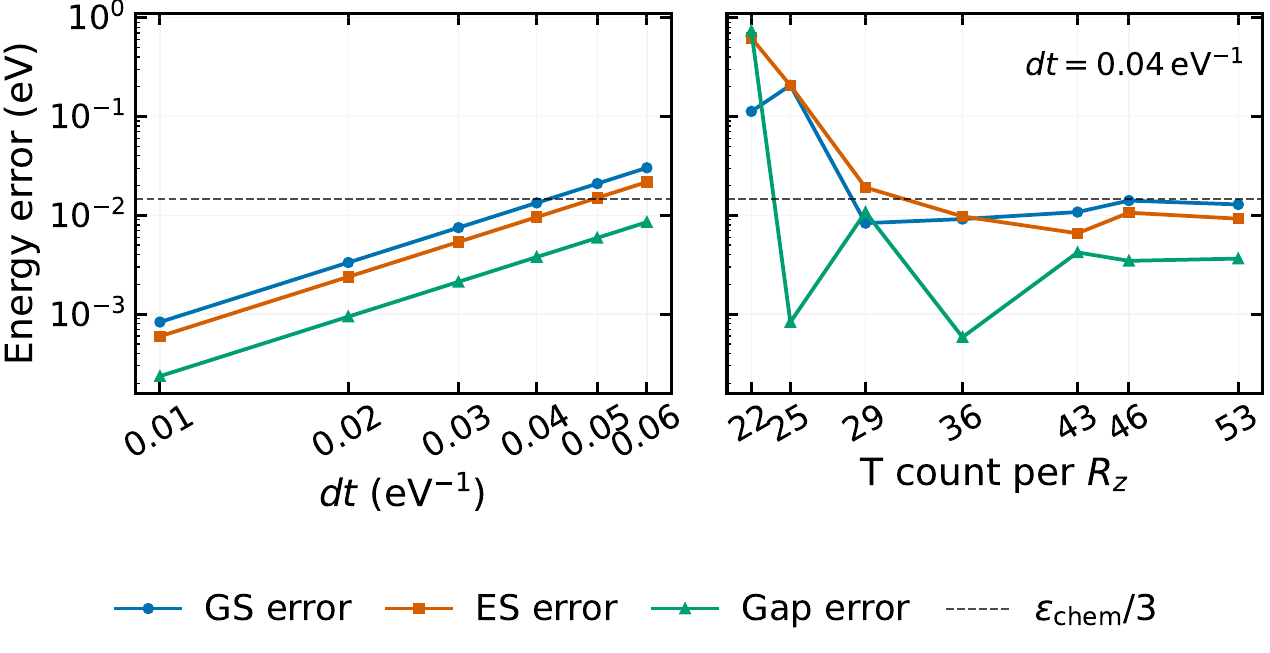}
\caption{Left) The Trotter energy eigenvalue error on the ground state, the first excited state, and the energy gap between those states of TMM - PPP unitary at different time steps. We require the eigenvalue error to be below $1/3$ of chemical accuracy to ensure sufficiently accurate eigenvalue estimation. ~\cite{Campbell_2021, baysmidt2026quantumsimulationnanographenestrotter} Right) The Trotter energy eigenvalue error with a fixed time step, $t = 0.04 \; \mathrm{eV}^{-1}$, as a function of the average T gate count per arbitrary rotation using gridsynth ~\cite{Gridsynth}.}
\label{fig:tmm-energy-error}
\end{figure}

We consider a small eigenvalue estimation problem, focusing on a small and interesting trial system: trimethylenemethane (TMM). TMM is the smallest non-Kekulé conjugated hydrocarbon and has the chemical structure $\mathrm{C_4} \mathrm{H_6}$. We use an effective model, the Pariser-Parr-Pople (PPP) model \cite{Pariser1953, Pople1953, fabian2025pppmodelminimal}, to describe the electronic structure of TMM, which reduces the electronic structure of TMM to an 8-qubit problem ~\cite{baysmidt2026quantumsimulationnanographenestrotter}. We provide the fermionic and qubit Hamiltonians of the TMM-PPP model (written in \texttt{Openfermion QubitOperator} and \texttt{FermionOperator} formats) in our GitHub repository \url{https://github.com/NQCP/NQCP-AA-git-zx-compiler}. 
Note that the TMM problem within this model is exactly solvable classically - the purpose of this example is not to tackle a classically hard problem, but rather to demonstrate how our compilation framework can be used in the case of a small fault-tolerant quantum algorithm. Despite this, the problem considered here can be naturally extended to classically challenging problems following Ref~\cite{baysmidt2026quantumsimulationnanographenestrotter}.

We focus on calculating energies of low-lying electronic states using either standard or statistical Trotterized phase estimation techniques. These algorithms require implementing an approximate controlled Hamiltonian simulation unitary, $U$, and applying $U$ a sufficient number of times, $r$. The time evolution operator of the TMM-PPP model is implemented using the tile Trotterization scheme of Ref~\cite{AndreasTileTrotter:D}. Concretely, we first construct a Clifford + $R_z(\theta)$ circuit that implements a directionally controlled version of $U$. In this circuit, we ignore rotation synthesis error which allows us to isolate the Trotter error and fix a time step. We show the energy eigenvalue errors on the ground state (gs) energy, the first excited state (es) energy, and the energy gap between gs and es coming from our approximate unitary at different time steps in Fig.~\ref{fig:tmm-energy-error}(left). Following Refs.~\cite{Campbell_2021, baysmidt2026quantumsimulationnanographenestrotter}, we require the energy eigenvalue Trotter error to be below one-third of chemical accuracy ($\epsilon_{\mathrm{chem}}$) to ensure sufficiently accurate energy estimates. We therefore fix the time step, $t = 0.04 \; \mathrm{eV}^{-1}$, and can use this as input to determine the number of required Trotter steps for quantum phase estimation; see, for example, Eq.~(C6) in Ref~\cite{baysmidt2026quantumsimulationnanographenestrotter}. For a fault-tolerant implementation of this circuit, we use gridsynth for arbitrary rotation synthesis ~\cite{Gridsynth}, and need to consider the combined unitary error of Trotterization and synthesis. In Fig~\ref{fig:tmm-energy-error}(right), we fix the time step at $t = 0.04 \; \mathrm{eV}^{-1}$ (fixed Trotter error) and vary the number of T gates used in our synthesis scheme, showing that $29$ $T$ gates per rotation is sufficient to ensure ground state and gap eigenvalue errors below chemical accuracy when considering both Trotter error and rotation synthesis error.

Using this unitary with fixed time-step ($t = 0.04 \; \mathrm{eV}^{-1}$) and a fixed number of $T$ gates per arbitrary rotation synthesis error as our starting point, we consider the resource requirements for three cases of phase estimation: 1) standard quantum phase estimation (QPE) to calculate the ground state energy, 2) statistical phase estimation (Stat-QPE) to calculate the ground state energy, and 3) statistical phase estimation to calculate the energy gap between the ground and the first excited state (Stat-QPE-Gap). We use an adaptive phase estimation scheme for QPE  ~\cite{Kivlichan2020ImprovedTrotterization}, which requires a single ancilla qubit on top of the 8 system qubits ~\cite{Kivlichan2020ImprovedTrotterization}. Using this scheme, we estimate that $r \sim 1680$ Trotter steps are sufficient for calculating the ground state energy up to chemical accuracy. Stat-QPE trades reduced circuit depths for increased sampling overhead, and previous results have shown that Stat-QPE circuit depths can be reduced by 1-2 orders of magnitude compared to QPE ~\cite{Ding2023, Ding2023simultaneous}. We estimate that the most expensive run of our Stat-QPE requires approximately $r \sim 40$ Trotter steps. Lastly, it was recently shown that calculating energy gaps between low-lying eigenstates of the type of model considered here is more efficient than absolute energy calculations ~\cite{baysmidt2026quantumsimulationnanographenestrotter}. Therefore, we estimate that the most expensive run of Stat-QPE-Gap requires around $r \sim 20$ Trotter steps (using a larger time-step than for the absolute energy estimation). Note that these resource estimates assume an initial state of sufficient overlap with the target state, and the cost of state preparation is not considered here. 

\subsection{Fourth order Trotter step of the Hubbard model \label{subapp:fourth-order-trotter-step-of-the-hubbard-model}}

We also consider the problem of performing a fourth-order Trotter step ($r=1$) of a $10 \times10$ square lattice Hubbard model, which is a 200-logical-qubit problem. We implement the fourth-order Trotter step using Qiskit's SuzukiTrotter product-formula class with order = 4 ~\cite{qiskit_suzukitrotter}. The arbitrary rotations are synthesized into $T$ gate sequences using gridsynth ~\cite{Gridsynth} with precision $10^{-10}$ following Ref.~\cite{khan2026architectingearlyfaulttolerant}. 
We implement the fourth-order Trotter step of a symmetry-shifted Fermi-Hubbard model according to ~\cite{Campbell_2021, AndreasTileTrotter:D}, which lowers the resource requirements compared to the naïve implementation, and choose $U/\tau = 8$ as our system parameter. 
Note that the rotation synthesis scheme employed here is largely independent of the rotation angle, which results in the T gate count being independent of the $U/\tau$ fraction and the magnitude of the time step. 

\section{Resource Estimation}
\label{app:resource-estimation}
In this section, we outline how resource estimation is performed for the four different architectures: the baseline and compact lattice-surgery layouts for local connectivity, the active-volume architecture for limited non-local connectivity, and the transversal active-volume architecture for effectively all-to-all connectivity. 
In App.~\ref{subapp:noise-model-and-distance-calculation}, we describe the noise models chosen for the different hardware types and show how the minimal required code distance can be calculated for each benchmark. Lastly, in App.~\ref{subapp:av-parallelization-and-bridge-qubit-demand}, we discuss the possibility of parallelization for the active-volume architectures.

\subsection{Noise model and distance calculation}
\label{subapp:noise-model-and-distance-calculation}
The logical error model is chosen according to the physical platform assumed for each architecture (see Table \ref{tab:platforms}).
For the lattice-surgery-based baseline, compact, and active-volume architectures, we use the standard circuit-level surface-code scaling form with an effective threshold scale near $1\%$~\cite{FowlerSurfFTQC,PBCGameSurface}. 
The logical error probability per logical tile per code cycle is approximated as
\begin{equation}
    p_L(p,d) = 0.1(100p)^{(d+1)/2}.
\label{circ-noise-model}
\end{equation}
For the neutral-atom t-AV architecture, we use an erasure-conversion scaling form with a higher effective threshold near $4\%$~\cite{ShrutiThresholdAtoms}:
\begin{equation}
    p_L^{\mathrm{NA}}(p,d) = 0.03(25p)^{(d+1)/2}.
\label{erasure-noise-model}
\end{equation}
We choose the minimum odd rotated-surface-code distance such that the probability of at least one logical failure during the computation is below the target algorithmic failure budget $\epsilon = 0.01$. 
For an architecture using $n_Q$ logical tiles for $n_C$ logical cycles, this is estimated by
\begin{equation}
    n_Q \cdot n_C \cdot d \cdot p_L(p,d) < \epsilon .
\end{equation}
Table~\ref{tab:min-distance} reports the resulting minimum distances at physical error rates $p=10^{-3}$ and $p=10^{-4}$ for each benchmark and compilation architecture.

\begin{table*}
\centering
\begin{tabular}{c l r r r r}
\toprule
Benchmark & Architecture & $n_Q$ & $n_C$
  & $d\,(p{=}10^{-3})$ & $d\,(p{=}10^{-4})$ \\
\midrule

\multirow{5}{*}{
\centering
\begin{tabular}{@{}c@{}}
\textbf{TMM} \\
(QPE to find absolute energies, Fig.\ref{fig:runtime-vs-count-qpe-abs}) \\
\textit{$N = 9$ logical qubits, $r$ = 1680}
\end{tabular}
}
& Baseline   & 32  & $1{,}370{,}880$       & 19 & 9 \\
& Compact    & 15   & $15{,}079{,}680$   & 21 & 11 \\
& Active volume
     & 50        & $895{,}440$ & 19 & 9 \\
& t-AV (photonics)
& $17.10$ & $342{,}720$ & 17 & 9 \\
& t-AV (atoms)
& $9.54$ & $1{,}365{,}840$ & 11 & 7 \\
\midrule

\multirow{5}{*}{
\centering
\begin{tabular}{@{}c@{}}
\textbf{TMM} \\
(Statistical QPE to find absolute energies, Fig. \ref{fig:runtime-vs-count-stat-qpe}) \\
\textit{$N = 9$ logical qubits, $r_{max}$ = 40}
\end{tabular}
}
& Baseline   & 32  & $32{,}640$       & 17 & 7 \\
& Compact    & 15   & $359{,}040$   & 17 & 9 \\
& Active volume
     & 50        & $21{,}320$ & 17 & 7 \\
& Transversal Active volume
& $9.54$ & $32{,}520$ & 15 & 7 \\
& t-av (atoms error model)
& $9.54$ & $32{,}520$ & 9 & 5 \\
\midrule

\multirow{5}{*}{
\centering
\begin{tabular}{@{}c@{}}
\textbf{TMM} \\
(Statistical QPE to find energy gaps, Fig. \ref{fig:runtime-vs-count-tmm}) \\
\textit{$N = 9$ logical qubits, $r_{max}$ = 20}
\end{tabular}
}
& Baseline   & 32  & $16{,}320$       & 15 & 7 \\
& Compact    & 15   & $179{,}520$   & 17 & 9 \\
& Active volume
     & 50        & $10{,}660$ & 15 & 7 \\
& Transversal Active volume
      & $17.10$  & $4{,}080$ & 13 & 7 \\
      & t-av (atoms error model)
      & $12.07$  & $8{,}140$ & 7 & 5 \\
\midrule

\multirow{5}{*}{
\centering
\begin{tabular}{@{}c@{}}
\textbf{Fermi-Hubbard} \\
(Precision Hamiltonian Simulation, Fig. \ref{fig:runtime-vs-count-fh}) \\
\textit{$N = 200$ logical qubits, $r$ = 1}
\end{tabular}
}
& Baseline      & 800   & $805{,}804$  & 23 & 11 \\
& Compact        & 303   & $8{,}863{,}844$ & 23 & 11 \\
& Active volume
    & 3304     & $142{,}399$ & 21 & 11 \\

& Transversal Active volume
& $209.43$ & $201{,}452$ & 19 & 9 \\
& t-av (atoms error model)
& $204.22$ & $402{,}903$ & 11 & 7 \\

\bottomrule
\end{tabular}

\caption{Minimum sufficient odd rotated-surface-code distance $d$ for each architecture variant
at two physical error rates, targeting the total logical error
$n_Q \cdot n_C \cdot d \cdot p_L(p,d) \leq 0.01$ on three TMM use cases (including phase estimation qubit) and one Fermi-Hubbard benchmark.
For Active-volume,
Transversal-Active-volume and t-av variants, the single entry shown is
selected by the lowest distance value; ties use the lower tile count. 
Tile count only includes logical tiles participating in computation; magic state factory tiles are excluded.}
\label{tab:min-distance}
\end{table*}

\subsection{Stat-QPE and QPE-Abs resource estimates}
\label{subsubapp:stat-qpe-resource-estimates}

\begin{figure*}[!t]
\includegraphics[width=\linewidth]{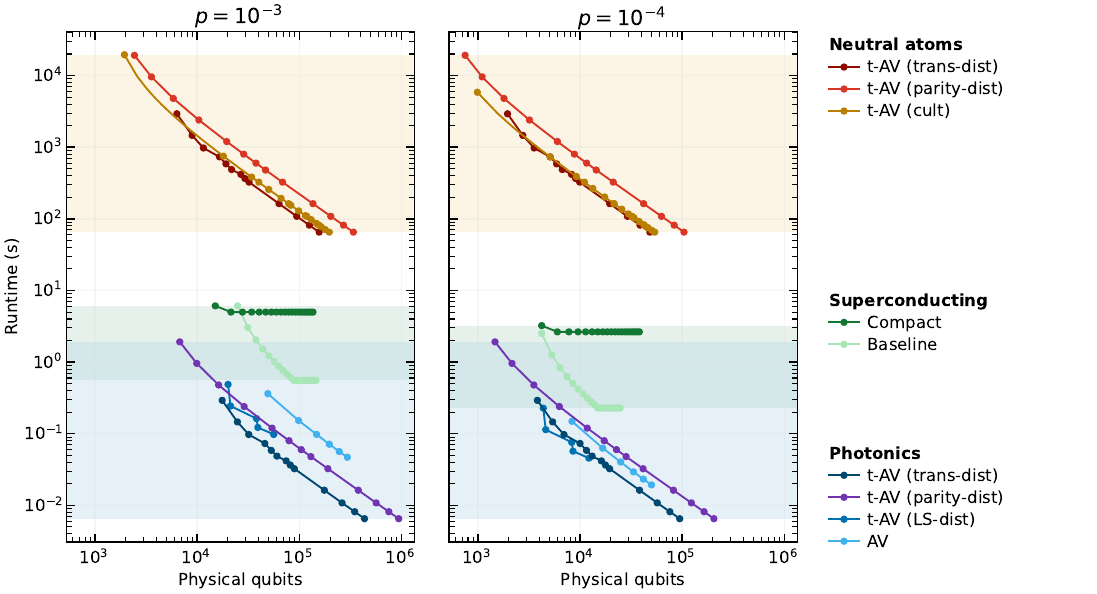}
\caption{Runtime vs.\ physical-qubit count for the Stat-QPE for TMM at error rates $p=10^{-3}$ (left) and $p=10^{-4}$ (right). See the caption of Fig.~\ref{fig:runtime-vs-count-tmm} for further details. Photonic t-AV reaches the lowest runtimes but at slightly higher qubit counts than neutral-atom t-AV. t-AV for neutral atoms achieves the smallest space footprint among all.}
\label{fig:runtime-vs-count-stat-qpe}
\end{figure*}

\begin{figure*}[!t]
\includegraphics[width=\linewidth]{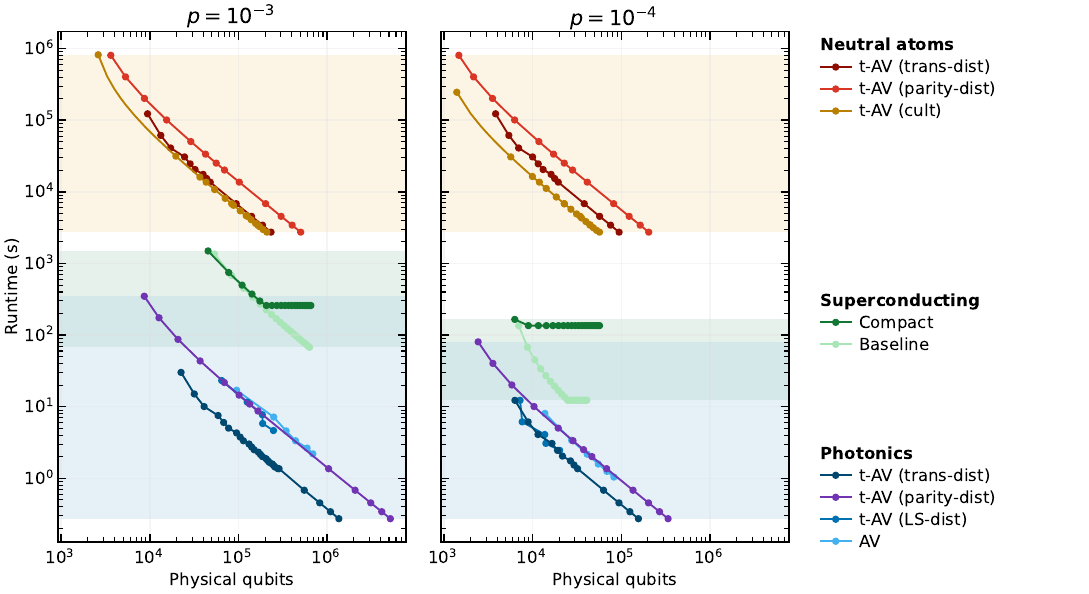}
\caption{Runtime vs. physical-qubit count for the QPE-Abs for TMM at error rates $p=10^{-3}$ (left) and $p=10^{-4}$ (right).  
See the caption of Fig.~\ref{fig:runtime-vs-count-tmm} for further details. Photonic t-AV reaches the lowest runtimes but at slightly higher qubit counts than neutral-atom t-AV. t-AV for neutral atoms achieves the smallest space-footprint among all}
\label{fig:runtime-vs-count-qpe-abs}
\end{figure*}

In this section, we continue the runtime vs qubit count analysis of Sec. 
\ref{sec:results} on Stat-QPE and Abs-QPE for our TMM benchmark as shown in Fig.~\ref{fig:runtime-vs-count-stat-qpe} and Fig.~\ref{fig:runtime-vs-count-qpe-abs}.

\textbf{TMM, Stat-QPE benchmark}
Stat-QPE requires the T state error rate to be below $3.1 \times 10^{-7}$. 
Although this is a tighter requirement than the Stat-QPE-Gap benchmark (Stat-QPE uses twice as many Trotter steps and therefore consumes roughly twice as many $T$ states), it is still comfortably met by 15-to-1 distillation; we therefore use the same distillation factories as for the Stat-QPE-Gap benchmark.
Cross-platform analysis follows a similar trend, but the runtime for Stat-QPE is larger than for Stat-QPE-Gap due to the doubled number of Trotter steps. 
The factor by which runtime is increased depends on whether the distillation cost scales with the code distance:
For the t-AV variants with constant-time distillation (parity-dist and trans-dist), the runtime is independent of the code distance, so the ratio is determined solely by the Trotter-step count and equals exactly $2$ at both error rates.
For cases whose distillation time scales with the code distance (Baseline, Compact, and AV architectures), whose runtime is proportional to $d$, the ratio can exceed $2$ when the two benchmarks require different code distances. 
This occurs only at the higher error rate $p=10^{-3}$. 
There, AV and Baseline both require $d=17$ for Stat-QPE versus $d=15$ for Stat-QPE-Gap, giving a ratio of $2.27$, and the t-AV lattice-surgery-factory variant gives $2.31$. 
Compact, by contrast, uses $d=17$ for both benchmarks and therefore remains at exactly $2.0$.
At $p=10^{-4}$, the two benchmarks share the same code distance in every case, so the ratio remains $2$ throughout.
In summary, for the distance-scaling cases, the runtime ratio is
\begin{equation}
\frac{t_{\text{Stat-QPE}}}{t_{\text{Stat-QPE-Gap}}} = \frac{r_{\text{Stat-QPE}}}{r_{\text{Stat-QPE-Gap}}}\times\frac{d_{\text{Stat-QPE}}}{d_{\text{Stat-QPE-Gap}}}
\label{eq: runtime-scaling-tmm}
\end{equation}
which reduces to $2$ whenever the two benchmarks share the same code distance.
In summary, Constant time distillation variants (t-AV (Trans-dist), t-AV (Parity-dist), t-AV (Cultivation)) and compact require 2x more time than the Stat-QPE-gap benchmark, AV and baseline take 2.27x more, and t-AV with LS-dist takes 2.31x more runtime. 

\textbf{TMM, Abs-QPE benchmark}
Abs-QPE benchmark requires the $T$-state error rate to be below $7.3\times10^{-9}$.
This is below the output error rate of a single-stage 15-to-1 factory at $p = 10^{-3}$, so wherever a single-stage factory cannot meet the requirement, we use the corresponding two-level distillation factories. 
For the Baseline, Compact, and AV architectures, a concatenated $(15\text{-to-}1)\times(8\text{-to-CCZ})$ factory is used at $p=10^{-3}$, whereas at $p=10^{-4}$, due to the lower physical error rate, a single factory is sufficient.
The runtime--qubit-count curves again follow a similar trend, but now require substantially more runtime owing to the much larger number of Trotter steps.
For the constant-time distillation cases (t-AV Parity-dist and Trans-dist, and Cultivation), the runtime is independent of the code distance, so the increase relative to Stat-QPE-Gap is exactly the Trotter-step ratio, $1680/20 = 84$, at both error rates.
For the distance-dependent factory cases (Baseline, Compact variant), whose runtime is proportional to $d$, the increase is given by Eq. \ref{eq: runtime-scaling-tmm}. 

\subsection{Parallelization in AV architecture and bridge qubit resource estimates}
\label{subapp:av-parallelization-and-bridge-qubit-demand}
\begin{figure}[tbp]
\includegraphics[width=\columnwidth]{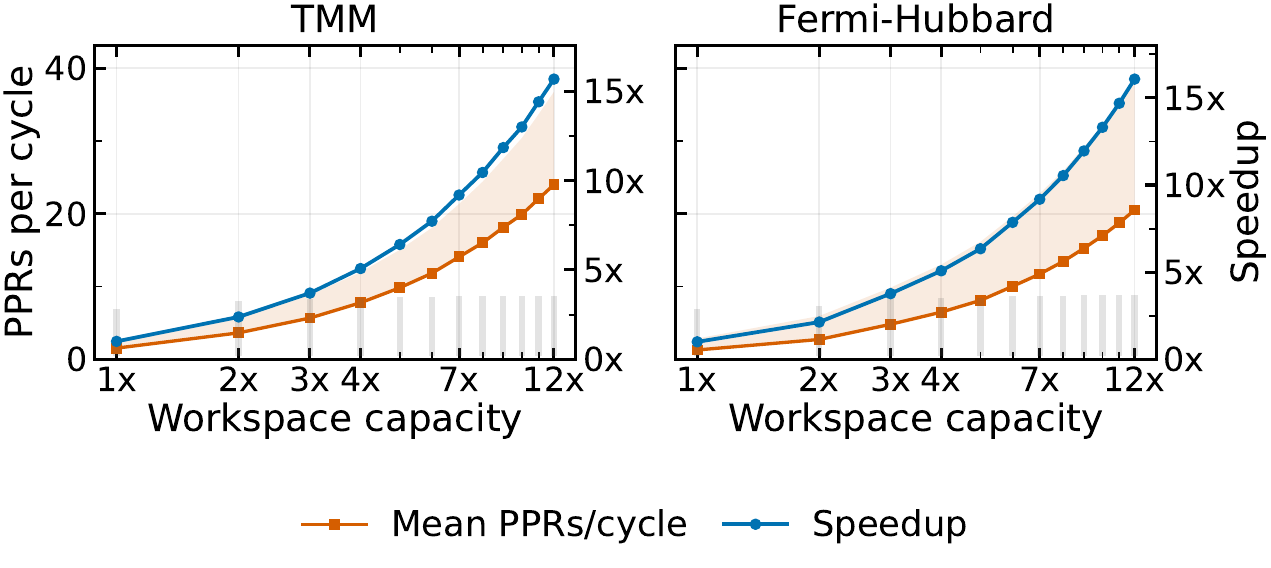}
\caption{PPR parallelization scaling in the AV architecture for TMM and Fermi-Hubbard
  (both per Trotter step), as a function of workspace capacity normalized to the
   minimum capacity needed to run the computation (1× = 25 logical qubits for TMM, 1× = 413
  logical qubits for Fermi-Hubbard). The orange curve (left axis) shows the mean
   number of PPRs scheduled per cycle, with the shaded orange band spanning one
  standard deviation above the mean; the blue curve (right axis) shows the
  Speedup in cycle count relative to the 1× baseline with bridge qubit parallelization. The gray vertical bars
  Indicate the mean workspace utilization at each capacity point, with taller bars
  indicating higher utilization. Both circuits achieve comparable speedup (~16×)
  at 12× capacity, with workspace utilization increasing from $73\%$ at 1× to $~96\%$
  at 12×.
  } 
  \label{fig:ppr-parallelization}
\end{figure}

In the AV architecture, parallelization of PPR depends on the workspace capacity. 
A minimum workspace capacity is set to the maximum number of logical blocks required among all the PPRs.
Increasing the workspace capacity allows more logical networks to be scheduled in parallel, reducing the total runtime in terms of logical cycles relative to an unparallelized schedule, as shown in Fig.~\ref{fig:ppr-parallelization}.
Increasing workspace capacity gives sublinear speedup in runtime, but workspace utilization (grey bars) saturates at $4$--$6\times$ for both benchmarks.
Therefore, a capacity around $4$--$6\times$ is enough to run the algorithm with optimal spacetime volume for both circuits.
The scheduler takes $17$--$33\%$ more PPRs per cycle for TMM, due to their low weight compared to the Fermi-Hubbard case.

\begin{figure}[tbp]
\centering
\includegraphics[width=1\columnwidth]{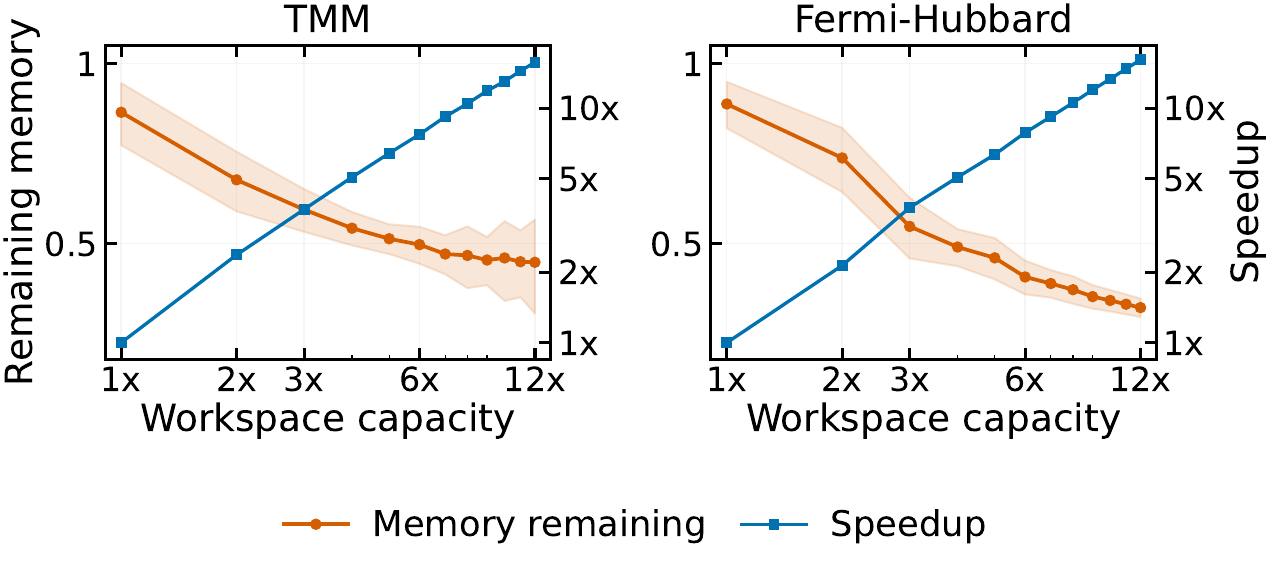}
\caption{Parallelization scaling for the AV architecture across workspace capacities
from 1× to 12× the minimum capacity needed to run the algorithm on an AV computer (1× = 25 logical blocks for TMM, 1× = 413 logical blocks for Fermi-Hubbard). 
The orange curve (left axis) shows the fraction of memory capacity remaining after bridge-qubit allocation, with the shaded band spanning ±1 standard deviation across cycles within the Trotter step. The blue curve (right axis) shows the speedup in logical cycle count relative to the 1× baseline. 
Both circuits achieve comparable speedup
scaling when capacity is normalized to their respective minimum workspace demand, but Fermi-Hubbard incurs substantially higher bridge-qubit demand per cycle at every capacity point, reflecting the denser qubit-sharing structure of its PPR network. }
\label{fig:bell pairs av}
\end{figure}

Higher parallelism can increase the bridge-qubit demand per cycle if the circuit contains many high-weight PPRs.
All bridge qubits live in memory; therefore, it is important to know the remaining capacity of memory to host other stale qubits from pending corrections. 
Figure~\ref{fig:bell pairs av} shows this tradeoff for TMM and Fermi-Hubbard.
TMM is small, so the runtime quickly saturates with workspace capacity; after some point, extra workspace gives little additional benefit.
Fermi-Hubbard is larger and benefits from a wider capacity range.
The memory-remaining curve indicates whether a given workspace capacity is bridge-resource constrained.
Low remaining memory means bridge-qubit demand is limiting parallel execution, whereas high remaining memory with little further runtime improvement indicates unused workspace capacity.
In this AV analysis, we assume the distillation block supplies enough $\ket{T}$ states to match the scheduled consumption rate.

\end{document}